%% file: bare_jrnl.tex
\documentclass[10pt,conference]{IEEEtran}
\usepackage{cite}
\usepackage{graphicx}
\usepackage{setspace,graphicx}%,subfigure
\usepackage{url}
\usepackage{xcolor}
\usepackage{amsmath,amssymb,amsfonts}

\usepackage[normalem]{ulem} %\sout

\usepackage{booktabs}
\usepackage{multirow}

\usepackage{hyperref}
\usepackage{subfigure}

\newcommand\han[1]{}

\newcommand{\rev}{}

\IEEEoverridecommandlockouts
\begin{document}
\title{\huge Time-Sensitive Networking (TSN) for Industrial Automation: Current Advances and Future Directions}

\author{
    \IEEEauthorblockN{
        Tianyu Zhang\IEEEauthorrefmark{2}$^*$\thanks{${}^*$The first two authors have equal contribution to this work.},
        Gang Wang\IEEEauthorrefmark{3},
        Chuanyu Xue\IEEEauthorrefmark{2},
        Jiachen Wang\IEEEauthorrefmark{2},
        Mark Nixon\IEEEauthorrefmark{3},
        Song Han\IEEEauthorrefmark{2}
    }

    \IEEEauthorblockA{
        \IEEEauthorrefmark{2}School of Computing, University of Connecticut
    }
    \IEEEauthorblockA{
        \IEEEauthorrefmark{2}Email: \{tianyu.zhang, chuanyu.xue, jiachen.wang, song.han\}@uconn.edu
    }
    \IEEEauthorblockA{
        \IEEEauthorrefmark{3}Emerson Automation Solutions
    }
    \IEEEauthorblockA{
        \IEEEauthorrefmark{3}Email: \{gang.wang2, mark.nixon\}@emerson.com
    }
}

\maketitle

\begin{abstract}
With the introduction of Cyber-Physical Systems (CPS) and Internet of Things (IoT) technologies, the automation industry is undergoing significant changes, particularly in improving production efficiency and reducing maintenance costs. Industrial automation applications often need to transmit time- and safety-critical data to closely monitor and control industrial processes. 
{\rev Several Ethernet-based fieldbus solutions, such as PROFINET IRT, EtherNet/IP, and EtherCAT, are widely used to ensure real-time communications in industrial automation systems. These solutions, however, commonly incorporate additional mechanisms to provide latency guarantees, making their interoperability a grand challenge.}
%There are a number of solutions to meet these requirements (e.g., priority-based real-time schedules). However, due to their different processing capabilities (e.g., in the end devices and network switches), different vendors may come out with distinct solutions, and this makes the large-scale integration of devices from different vendors difficult or impossible. 
The IEEE 802.1 Time-Sensitive Networking (TSN) task group was formed to enhance and optimize IEEE 802.1 network standards, particularly for Ethernet-based networks. These solutions can be evolved and adapted for cross-industry scenarios, such as large-scale distributed industrial plants requiring multiple industrial entities to work collaboratively. This paper provides a comprehensive review of current advances in TSN standards for industrial automation. It presents the state-of-the-art IEEE TSN standards and discusses the opportunities and challenges of integrating TSN into the automation industry. Some promising research directions are also highlighted for applying TSN technologies to industrial automation applications.
\end{abstract}

%\markboth{IEEE COMMUNICATIONS SURVEYS & TUTORIALS}%
%{TSN for Industrial Automation}
%{Wang \MakeLowercase{\textit{et al.}}: Time-Sensitive Networks for Industrial Automation}

%\keywords{Time-Sensitive Networking (TSN), IEEE 802.1Q, Industrial Automation}

%\IEEEpeerreviewmaketitle

%\input{sec/notes.tex}
\input{sec/intro.tex}

\input{sec/Preliminary.tex}
\input{sec/TSN.tex}
\input{sec/Opportunities.tex}
\input{sec/Challenges.tex}

\input{sec/Vision.tex}
\input{sec/conclusion.tex}

\bibliographystyle{IEEEtran}
\bibliography{refs}

\end{document}

%% file: sec/intro.tex
\section{Introduction}
\label{Sec:Intro}
Industrial automation systems commonly employ a hierarchical architecture to perform designed control and automation processes~\cite{mahmoud2019architecture}. Ethernet-based fieldbus communication systems are currently dominating the automation industry, with multiple protocols and standards available~\cite{thomesse2005fieldbus}. However, different vendors may select different industrial Ethernet protocols for use in their devices, resulting in incompatibilities among the deployed equipment. This phenomenon contributes to industrial automation architectures being hierarchical, custom-built, and inflexible when integrating devices from different vendors or standards~\cite{leitao2016industrial}. Fortunately, driven by the recent advances in Industrial Internet of Things (IIoT) technologies, many technical initiatives are pushing industrial automation applications to be more flexible, interoperable, and seamless~\cite{gilchrist2016industry}. One of the most important requirements for industrial automation is real-time and deterministic communication, which is essential for realizing mission-critical control processes~\cite{IntelWP}. 

Critical traffic flows generated by industrial automation applications require bounded low latency and low jitter to improve production efficiency and reduce communication costs. Typically, these critical traffic flows need to share the communication medium (e.g., Ethernet) with non-critical flows (e.g., those with less severe timing constraints) originating from the same applications. Under these conditions, it is imperative to guarantee the timing behavior of critical traffic and provide temporal isolation from non-critical communications. %In general, industrial automation is driven by an accelerated integration of Information Technology (IT) (e.g., Ethernet, TCP/IP~\cite{comer1991internetworking}) and Operations Technology (OT) (e.g., Embedded Systems and Cyber-Physical Systems (CPS)~\cite{lee2008cyber})~\cite{leitao2016industrial}. And, the IEEE 802 is traditionally defined only for the protocols of information communication (IT domain), and now it is targeting on defining the protocols for IT-based standards for embedded system communication (a branch of OT domain)~\cite{steiner2016next}. Especially, 
{\rev The IEEE 802.1 Time-Sensitive Networking Task Group (TSN TG), evolved from the former IEEE 802.1 Audio Video Bridging (AVB) TG}, addresses this need by designing general-purpose protocols applicable to various fields, such as factory automation, process automation, substation control, {\rev and aerospace applications.}

The IEEE TSN TG currently aims to improve the reliability and real-time capabilities of the Ethernet standard (e.g., IEEE 802.3~\cite{7428776} and IEEE 802.1D~\cite{1309630}). It focuses on several essential aspects of the IEEE AVB standards crucial for industrial automation, including reduced latency, deterministic transmission, independence from physical transmission rates, fault tolerance without additional hardware, and interoperability of solutions from different vendors. 
{\rev Compared to traditional Ethernet-based fieldbus systems, the advantage of TSN is also manifold, including vendor neutrality, higher throughput, more network configuration flexibility, and better scalability~\cite{pfrommer2018open}.} 

TSN is a collection of standards, standard amendments, and projects published or under development by the TSN TG within the IEEE 802.1 Working Group (WG)~\cite{finn2018introduction}. There are four main pillars on which TSN is built: 1) time synchronization, 2) guaranteed end-to-end (e2e) latency, 3) reliability, and 4) resource management. These characteristics make TSN a strong candidate for meeting special requirements in industrial automation, such as deterministic communication, ultra-low communication latency and extremely high reliability. 
{\rev While TSN standardization efforts are ongoing, several manufacturers have already demonstrated the promising performance of TSN, showing much higher determinism than current state-of-the-art solutions~\cite{maxim2017delay,arestova2021simulative}.} However, the benefits of TSN come with challenges that need to be addressed in the deployment of of industrial automation systems. These challenges include stringent requirements on network synchronization precision, increased traffic scheduling complexity, integration with wireless devices, etc. 

This paper provides a comprehensive review of the current advances in standardization and research efforts related to TSN for industrial automation. We first give a systematical introduction to the published TSN standards relevant to industrial automation systems and explore the challenges each standard attempts to address. We then highlight how and to what extent these standardization efforts empower Ethernet applications, supporting the new requirements raised by current and future industrial use cases. 
{\rev Note that, in addition to the automation industry, deploying TSN technologies is of great interest in many other industries requiring deterministic, low-latency, and high-reliability communications, including automotive applications~\cite{ashjaei2021time}, aerospace~\cite{fiori2024proposal}, and healthcare~\cite{lu2023time}, which are not the focus of this survey. }

The rest of this article is organized as follows. Section~\ref{Sec:BG} provides the background of industrial automation and IEEE TSN technologies. Section~\ref{Sec:Sta} describes the up-to-date TSN standardization efforts in detail, and Section~\ref{Sec:oppo} discusses the integration of TSN into industrial automation systems. Section~\ref{Sec:Chal} discusses the challenges in each category of TSN standards. Section~\ref{Sec:Visi} presents the future directions related to TSN R\&D, and Section~\ref{Sec:Concl} concludes the article.

%% file: sec/Preliminary.tex
\section{Background}\label{Sec:BG}
With the introduction of CPS and IoT technologies, {\rev the automation industry is undergoing tremendous changes in architecture design and system development.} These recent technological advancements enable the interconnection of industrial assets on a broader and more fine-grained scale~\cite{wollschlaeger2017future}. In this section, we provide the background information of industrial automation and TSN technologies.

\subsection{Industrial Automation}\label{Sec:IA}
Industrial automation is an industry concept that utilizes various sensors, actuators, robotic devices, control systems, {\rev and information technology (IT) systems} to connect and manage different processes and machinery across multiple industries, replacing operations originally performed by humans~\cite{bahrin2016industry}. 

\subsubsection{Recent Trends in Industrial Automation}\label{sssec:trend}

%\begin{figure*}
%  \centering
%  \includegraphics[width=18cm]{Figs/IAS-Comm.png}
%  \caption{The milestones in the evolution of industrial communication and related technology fields~\cite{wollschlaeger2017future}. 2G: second generation; 3G: third generation; 4G: fourth generation; arPaNET: advanced research Projects agency Network;  GSm: global system for mobile communication; ISO: International Organization for Standardization; LTE: long-term evolution;  WLaN: wireless local area network; WWW: World Wide Web; aSi: actuator/sensor interface; EIB: European installation bus; caN:  controller area network; PrOFIBUS: process field bus; FIP: factory instrumentation protocol; harT: highway addressable remote transducer;  hSPa: high-speed packet access; LIN: local interconnect network; LON: local operating network; maP: manufacturing automation protocol;  mmS: manufacturing messaging specification; PrOWaY: process data highway; SOaP: simple object access protocol; TcP: transport control protocol;  TTP: trime triggered protocol; UmTS: universal mobile telecommunications system; UWB: ultrawide band; SdS: smart distributed system; PrOFINET: process field net;  EthercaT: Ethernet for control automation technology. }
%  \label{Fig:IASComm}
%\end{figure*}

% https://slideplayer.com/slide/5905500/

The industry has undergone three revolutions: mechanization, electrification, and information. The fourth industrial revolution (also referred to as “Industry 4.0”), currently underway, is marked by the pervasive deployment of IoT devices and services. In this revolution, a wide range of devices are being deployed in a self-organizing manner, typically relying on control and communication systems to manage their operation and interaction~\cite{danielis2014survey}. 
For example, in Supervisory Control and Data Acquisition (SCADA) systems~\cite{boyer2009scada}, proprietary communication systems have been mostly replaced by Sensorbus~\cite{ribeiro2005sensorbus} and fieldbus systems~\cite{thomesse2005fieldbus}. %Additionally, due to mass production and low component prices, industrial domains commonly adopt Ethernet as the communication medium. Meanwhile, numerous academic research efforts are optimizing Ethernet-based communication systems to meet the stringent requirements of industrial applications, such as in-vehicle networking in automotive Ethernet~\cite{matheus2017automotive}, Avionics Full Duplex Switched Ethernet (AFDX)~\cite{bauer2010improving} in avionics, and {\rev TTE/FTTE~\cite{eramo2022flexible,eramo2018definition} in aerospace applications}. 

\begin{figure}[tb]
  \centering
  \includegraphics[width=0.7\columnwidth]{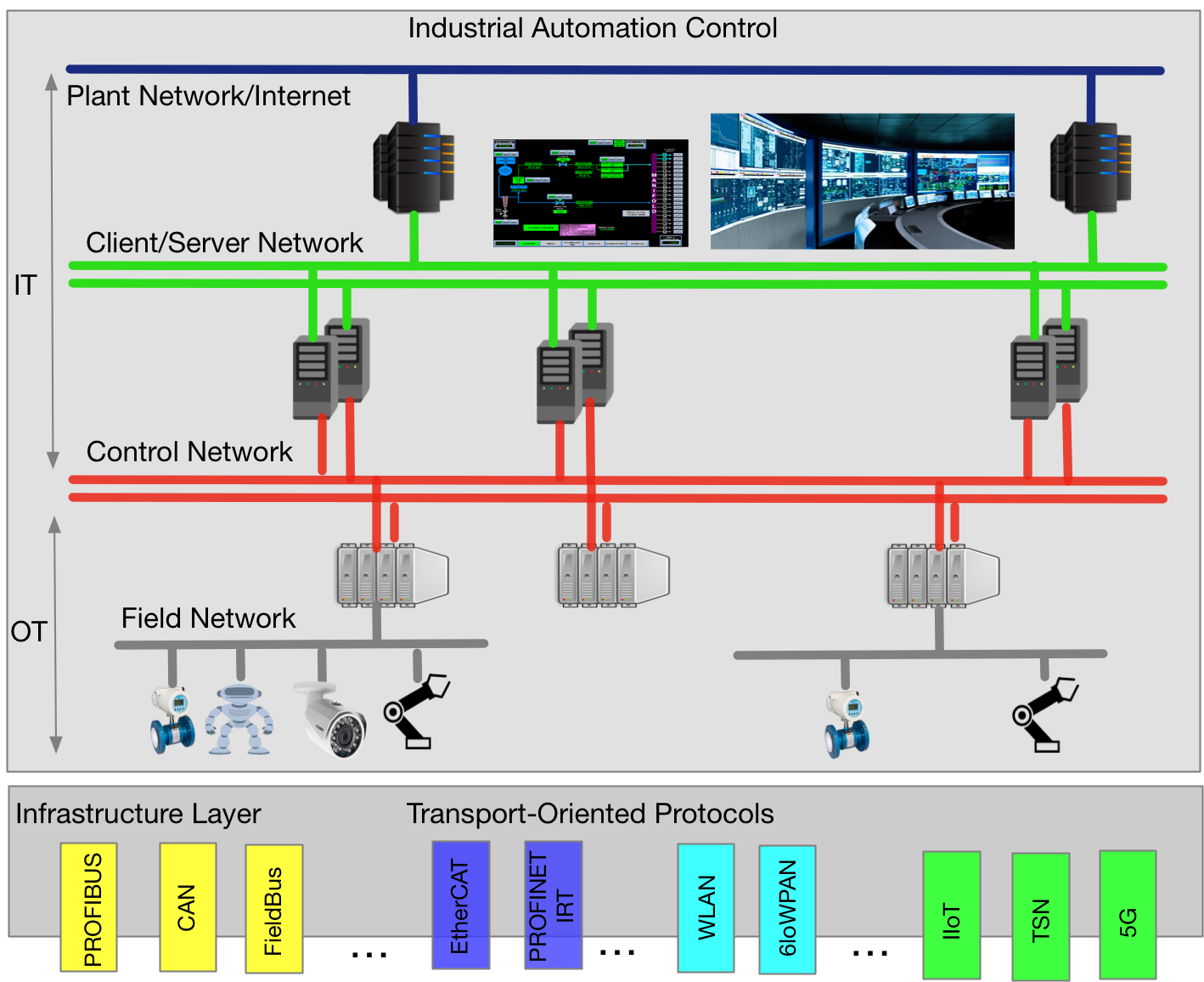}
  \caption{Example of industrial automation control hierarchy which consists of IT and OT parts. }
  \label{Fig:IAControl}
  %\vspace{-0.15in}
\end{figure}

{\rev The Industry 4.0 revolution posts significantly different requirements on industrial automation systems design. For example, the Industrial IoT (IIoT) paradigm advocates for a flat cloud of interconnected devices rather than a complex hierarchy. This shift necessitates a more unified communication system based on IP across all functional layers, where typical requirements on industrial automation systems such as time synchronization, low latency, determinism, and convergence must be met~\cite{bruckner2019introduction}. A flatter hierarchy also demands robust communication systems that support the coexistence of information technology (IT) and operational technology (OT) systems in industrial automation. Fig.~\ref{Fig:IAControl} illustrates an example of the industrial automation control hierarchy comprising IT and OT components, where IT technologies focus on network connectivity and data communication, whereas OT technologies focus on process operation and the control of field devices~\cite{boyes2018industrial}. The infrastructure layer provides various transport-oriented protocols to interconnect different IT and OT components. 
}

%{\rev Industrial automation typically involves both information technology (IT) and operation technology (OT) systems. Fig.~\ref{Fig:IAControl} shows an example of the industrial automation control hierarchy, which consists of IT and OT parts. IT focuses on network connectivity and data communication, while OT focuses on operational processes and the control of field devices~\cite{boyes2018industrial}. The infrastructure layer provides various transport-oriented protocols to interconnect different components. }
Industrial automation encompasses a variety of systems, including continuous condition monitoring systems, industrial control systems, and prevention/protection systems. While the functional requirements for different automation systems may vary across domains, they share similarities in terms of physical and logical organization complexity. Additionally, they share common requirements for determinism, reliability, interoperability, and traffic convergence~\cite{breivold2015internet}. 

%\vspace{0.05in}
$\bullet$ Timing and determinism: Industrial automation typically runs real-time applications with stringent requirements on their temporal behavior and accuracy when responding to internal and external events~\cite{sisinni2018industrial}. Beyond network throughput, the commonly used performance metrics, packet transmission latency and its time variations (jitter) are critical concerns for many industrial control systems~\cite{zhang2018distributed}. Timing interactions can complicate different procedures. For example, in a switched Ethernet network, achieving deterministic delay is challenging due to the presence of skew or drift in timing signal frames. In addition, the transmission of Ethernet frames can be delayed if the output port on a switching device is busy. These factors accumulate non-deterministic delays in data transmission, which are unsuitable for real-time industrial applications~\cite{zurawski2014industrial}. 
Therefore, to ensure correct operation, industrial automation systems require a certain degree of determinism~\cite{wollschlaeger2017future}. 

%\vspace{0.05in}
$\bullet$ Reliability and availability: Production losses in industrial automation due to unexpected stops caused by failure or deterioration of the communication environment are unacceptable. %A prevalent practice to avoid these unplanned stops is to ensure stable transmission conditions. Given the trend of decentralization in many applications, a robust industrial communication system is essential, particularly in process automation. Maintaining reliability and accessibility is key to ensuring communication quality~\cite{gong2019reliable}.
{\rev Thus, the reliability and availability\footnote{{\rev Reliability and availability are two similar concepts in the context of industrial automation with slight differences. Availability not only takes the possibility of failure but also the possibility of repair into account~\cite{bello2019perspective}.}} of the system are critically important due to the need for accurate and continuous operation in any condition. Reliability can be quantified using appropriate measures such as mean time between failures, or the probability of no failure within a specified period of time~\cite{gong2019reliable}. Many mission-critical industrial applications often aim for an uptime on the order of 99.999\% (known as "five nines" reliability), e.g., 99.9\% to 99.9999\% for closed-loop control~\cite{cavalcanti2020wireless}.
}

%\vspace{0.05in}
$\bullet$ {\rev Interoperability: An industrial automation system typically consists of diverse devices interconnected through varied technologies.  This heterogeneous system architecture necessitates the ability of disparate systems to communicate and share information or resources with one another, known as interoperability. Interoperability is crucial for industrial automation due to its many advantages. For example. by enabling seamless communication and coordination between various systems, businesses can experience enhanced accuracy and productivity. Real-time data exchange and coordinated control across the entire automation system also facilitate efficient decision-making, reducing errors and delays. Interoperability also improves scalability and flexibility, allowing for easier system expansion and modification. 
}
%Managing this wide range of components from different manufacturers is a challenging task. These components are intended to be incorporated into many different systems and the number of possible configurations is huge. 
%Due to the multitude and diversity of these integrated components, integration and engineering efforts to achieve interoperability are costly. These challenges have inspired many research efforts to overcome interoperability in industrial automation. Throughout the automation hierarchy, for example, OPC Unified Architecture (OPC UA) can address interoperability between components~\cite{henssen2014interoperability}. 

%\vspace{0.05in}
$\bullet$ {\rev Traffic convergence: Industrial automation applications make use of different traffic types for different functionalities, e.g. sensing, control, alarming, etc. The diverse traffic types have different characteristics and thus impose varied QoS requirements. These traffic can generally be classified into \textit{critical traffic} and \textit{best-effort} traffic. Critical traffic typically has stringent QoS requirements and different types of critical traffic may have particular QoS demands, depending on the specific application scenarios. IEC/IEEE 60802 group summarizes the traffic types for industrial automation (see Table.~1 in~\cite{60802}). Characteristics of these traffic types include deadline and latency, synchronization, transmission period, data size, and interference tolerance.  For example, isochronous control loops must meet guaranteed deadline requirements ($<$ 2 $ms$) and cannot tolerate packet loss. While cyclic traffic has more relaxed latency requirements (2 - 20 $ms$) and can tolerate some packet loss (1-4 frames)~\cite{ademaj2019industrial}.  In contrast to critical traffic, best-effort traffic generally does not have specific QoS requirements in any of these aspects.
}

{\rev To sum up, industrial automation applications have stringent and specific needs that revolve around ensuring real-time and deterministic communication, high reliability and availability, and interoperability to ensure the efficient, reliable, and safe operation of manufacturing processes and control systems, while supporting diverse traffic types. Among these needs, the requirement on deterministic real-time communication, typically evaluated using latency and jitter, plays the most critical role in industrial applications, which we will discuss further below.
}

%To enable robust industrial automation, there exist other critical requirements, e.g., scalability, fault tolerance, and security. Specific techniques used for networking in industrial automation are still undergoing changes. What hasn't changed over time, however, is the need to satisfy the fundamental requirements of the industrial domain such as hard time constraints, isochronic communication, low jitter, high reliability, and low cost.

\subsubsection{Deterministic Real-Time Communication}
Packet latency, one of the most critical QoS characteristics, typically refers to an end-to-end (e2e) packet delay from the moment when the sender initiates the transmission to its complete reception by the receiver. The requirement for low latency generally implies that the transmission time must be very short, often within milliseconds, to meet the necessary QoS requirements.  
Additionally, low-latency applications usually demand deterministic latency~\cite{zhang2017distributed}. For instance, to ensure the proper functioning of industrial automation systems, all frames within a specified application traffic flow must adhere to a pre-defined latency bound~\cite{qian2017xpresseth,zhang2018fd}. 
Some industrial applications also require probabilistic latency. For example, a pre-defined delay bound should be met with high probability, such as in multimedia streaming systems~\cite{imtiaz2009performance}, where occasional delay bound violations have negligible effect on perceived multimedia quality~\cite{hei2007inferring}.

Latency jitter, or jitter for short, refers to variations in packet latency. {\rev Industrial automation systems typically require very low jitter to ensure highly predictable and reliable communication, which is crucial for the proper functioning of industrial processes. Minimizing jitter is essential for maintaining the synchronization and timing precision needed for industrial applications, particularly in motion control, where low jitter is critical for controlling actuation devices.  Other industrial applications with low jitter requirements include but are not limited to machine tools (100 $ns$), automotive radar (20 $ns$), and professional audio (10 $ns$)~\cite{tiwhite}. }

{\rev Latency and jitter are the primary QoS metrics for industrial automation networks. When both packet latency and jitter can be bounded, the communication is considered deterministic, meaning that the message will be transmitted within a specified and predictable time frame. Determinism ensures that communication or output will not only be correct but also occur within a defined period. Industrial automation networks are typically deterministic, catering to many applications requiring such services, including condition monitoring~\cite{wang2019toward,windmann2016evaluation}, process automation~\cite{groover2007automation,mehta2014industrial}, smart manufacturing~\cite{fortis2015automatic}, printing machines~\cite{chen2015water}, and connected cars~\cite{berdigh2018connected,fallgren2018fifth}.}
%\\
%$\cdot$ Condition Monitoring~\cite{wang2019toward}~\cite{windmann2016evaluation}
%\\
%$\cdot$ Process Automation~\cite{groover2007automation}~\cite{mehta2014industrial}
%%\\
%%$\cdot$ Machine Tool~\cite{flynn2016hybrid}
%\\
%$\cdot$ Smart Manufacturing~\cite{fortis2015automatic}
%\\
%$\cdot$ Printing Machines~\cite{chen2015water}
%\\
%$\cdot$ Connected Cars~\cite{fallgren2018fifth}

%The performance of these applications is given by a set of indicators, such as cycle time, a number of communicating devices, and the synchronization accuracy of their clocks, payload per devices, etc. The typical timing constraints of these applications are illustrated in the Table~\ref{Tab:Timing}. % see the citation.

%The future of industrial Ethernet @ IEEE Magazine

\subsubsection{The Future of Industrial Ethernet}
% Survey on Real-Time Communication Via Ethernet in Industrial Automation Environments

%\begin{figure*}
%  \centering
%  \includegraphics[width=13cm]{Figs/RTE.png}
%  \caption{The classification scheme for real-time Ethernet}
%  \label{Fig:RTE}
%\end{figure*}

Currently, Ethernet-based fieldbus systems are prevalent for industrial automation using the widespread Ethernet technology. The implementation of Ethernet to connect field devices offers significant advantages as Ethernet allows for consistent integration at all levels of the hierarchy. In particular, Ethernet enables the vertical and horizontal integration of the industrial automation system from the field level to the application level~\cite{sauter2005integration}, which is essential for realizing the vision of IIoT. 
To achieve the required higher quality of data transmission, Real-Time Ethernet (RTE)~\cite{felser2005real} has become a standard in the automation industry today. However, there is no single standard at present but many different mutually incompatible implementations. Existing RTE solutions can generally be organized into three classes~\cite{wollschlaeger2017future}.

%\vspace{0.05in}
$\bullet$ Class A: Real-time services with cycle times ranging from 100 $ms$. Example implementations include Modbus-Interface for Distributed Automation (IDA), Ethernet/Industrial Protocol (IP), and Foundation Fieldbus (FF) high-speed Ethernet. This class builds on the entire TCP/IP transportation control suite and uses best-effort bridging.

%\vspace{0.05in}
$\bullet$ Class B: Real-time services are performed directly at the top of the Media Access Control (MAC) layer using approaches such as prioritization and Virtual Local Area Network (VLAN) targeting to separate real-time traffic from the best-effort traffic. For example, using Fast Ethernet~\cite{varadarajan1998ethereal}, the achievable cycle time is within 10 $ms$.

%\vspace{0.05in}
$\bullet$ Class C: {\rev Real-time communication is achieved by modifications of the Ethernet MAC layer, including strict traffic scheduling and high-precision clock synchronization. The achievable cycle time can be less than 1 $ms$. Some examples of implementations are EtherCAT~\cite{prytz2008performance}, Time-Triggered Ethernet (TTE)~\cite{steiner2009ttethernet} and its variation, Flexible Time-Triggered Ethernet (FTTE)~\cite{eramo2023max}. 
}
%Ethernet-based protocols are developed and adopted in real-time safety-critical domains~\cite{zhang2014task}. Besides the use cases in industry automation (e.g., EtherCAT, fieldbus), these features can be extended to actual deployments from non-industrial areas, e.g., AFDX~\cite{bauer2010improving} in avionics. 

{\rev Class C is the most potent class for meeting industrial automation requirements, particularly for TTE and FTTE, which enable determinism in the bandwidth and latency of Ethernet. However, these standards have distinct differences in their support for traffic heterogeneity, time-schedule traffic, time synchronization, and adherence to open standards, thus catering to slightly different needs and markets\footnote{{\rev Further discussion and comparison of TTE, FTTE and TSN can be referred to~\cite{zhao2018comparison,eramo2023performance}. This paper primarily focuses on TSN for industrial automation.}}. Furthermore, as pointed out by~\cite{danielis2018real}, which summarizes the requirements of industrial applications into R1 - R7, no industry-established Ethernet-based fieldbus technology can meet all these requirements. Some quantitative performance comparison results among several real-time Ethernet protocols can be found in~\cite{Schlegel2017,bruckner2018opc}. 
}
%Concurrently, standard Ethernet is evolving towards a real-time communication system. 

{\rev Meanwhile, standard Ethernet is evolving towards a real-time communication system that can be applied in industrial applications.} The IEEE TSN TG is working on improving the reliability and real-time capabilities of Ethernet standards. Specially, the task group addresses several critical shortcomings of the AVB standard, which are vital for industrial automation. These improvements include decreased latency and precise determinism, independence from physical transmission rates, fault tolerance without additional equipment, higher safety and security support, and interoperability among products from different manufacturers. In the following sections, we will detail each of these aspects. 

%{\rev In addition to IEEE TSN, there exist other standardization efforts to implement real-time Ethernet, e.g., TTEthernet (TTE)~\cite{steiner2009ttethernet} and Flexible Time-Triggered Ethernet (FTTE)~\cite{eramo2023max}, to provide guaranteed QoS performance for real-time applications. While all these standards enable determinism to bandwidth and latency of Ethernet, they have distinct differences in terms of heterogenous traffic support, time-schedule traffic, time-synchronization, and open standards, and thus, they cater to slightly different needs and markets. Further discussion and comparison of TTE, FTTE and TSN can be referred to~\cite{zhao2018comparison,eramo2023performance}. This paper, however, primarily focuses on TSN for industrial automation.}

\subsection{IEEE 802.1 Overview}
TSN is a recent IEEE 802.1 standardization effort aiming at integrating real-time capabilities in the Ethernet standard, which could eventually replace current industrial communication systems. Before delving into the details of the TSN standards, we first provide an overview of the IEEE 802.1 standard. The IEEE 802.1 standard is primarily based on the IEEE 802.1Q standard~\cite{8403927}, known as ``Bridges and Bridged Networks". Various standardization projects within different task groups modify and extend IEEE 802.1Q, incorporating amendments such as the latest version, IEEE 802.1Q-2018. These amendments, denoted as IEEE 802.1Qxx, indicate specific changes to the previous IEEE 802.1Q version.

The IEEE 802.1Q standard encompasses two main concepts: bridges and traffic. In an IEEE 802.1-enabled network, a bridge typically refers to a network entity that meets the mandatory or recommended functionalities specified by the corresponding standards. The IEEE 802.1Q standard defines specifications for both Local Area Network (LAN) and Wide Local Area Network (WLAN) bridging infrastructures. For instance, IEEE 802.1Q outlines the specifications for intra- and inter-communication for VLAN-aware bridges and bridged LAN networks, and it also specifies the corresponding protocol layers above the MAC and LLC layers~\cite{nasrallah2018ultra}. 
%We are interested in a special type of bridge in the course of TSN, the VLAN bridge as defined in clause 5.4 of 802.1Q, which has some subclauses (referring to clause 8) to define mandatory frame relaying and filtering functionality. IEEE 802.1Q describes frame forwarding duties within the Clause 8 bridge.

\begin{table}[]
\small
\centering
\caption{IEEE 802.1 Traffic Classes~\cite{8403927}: 0 represents the lowest priority and 7 represents the highest priority.}
\begin{tabular}{|c|l|}
\hline
%\rowcolor[HTML]{EFEFEF} 
\textbf{Priority} & \textbf{Traffic Class Description}                     \\ \hline
0        & Background information                        \\ \hline
1        & Best effort traffic                           \\ \hline
2        & Excellent effort traffic                      \\ \hline
3        & Critical application traffic                  \\ \hline
4        & “Video” stream with \textless{}100 $ms$ latency \\ \hline
5        & “Voice” stream with \textless{}10 $ms$ latency  \\ \hline
6        & Network control traffic, e.g., BGP, RIP       \\ \hline
7        & Critical control data traffic (CDT)           \\ \hline
\end{tabular}
\label{Tab:traffic}
%\vspace{-0.15in}
\end{table}

To achieve low latency for certain types of traffic in IEEE 802.1Q, priority-based traffic classes are defined to characterize the behavior of different traffic types, such as Class of Service (CoS)~\cite{802.1Q-2014}. The 802.1Q standard defines a queuing model for the transmission port, where up to eight traffic class queues can be implemented for each port. These queues have port-local priorities, as summarized in Table~\ref{Tab:traffic}, with the lowest priority zero and the highest seven~\cite{lim2012performance}.

A typical Ethernet-switched network consists of end stations and bridges (or switches)\footnote{{\rev In this paper, we use switch and bridge interchangeably without specified.}}. End stations typically provide the application data, while bridges relay the data between end stations, from the data source to the data destination. The duties of bridges can be broadly classified into traffic switching, traffic shaping, and traffic policing~\cite{bello2019perspective}. Applying IEEE 802.1 to industrial automation necessitates several new features, such as worst-case e2e latency, high efficiency, bandwidth isolation, and zero congestion loss between bridges and end stations.

There are two types of traffic-switching operations in bridged Ethernet networks: store-and-forward, and cut-through~\cite{duato1996necessary}. In the store-and-forward method, an Ethernet frame must be received entirely before sending it out from any transmission port. Conversely, the cut-through approach allows the transmission of an Ethernet frame to begin even before the bridge has fully received the frame. Note that when combining cut-through with deterministic behavior, the cut-through approach forwards frames without first checking their correctness. As a result, invalid frames (e.g., attributed to another stream) can still alter the state of the shaping algorithm and impact later valid frames. Due to these considerations, IEEE 802.1Q does not specify whether the cut-through mode is legitimate in a bridge, as some standards do not support cut-through operations. In addition to industrial automation, there are many use cases where cut-through is essential to meet exceptionally low latency demands~\cite{koulamas2004using}. %; there is also ongoing discussion about whether to explicitly include cut-throughs and formulate solutions. 

%%%%%Due to its simple connection mechanisms and protocol operations, Ethernet has been widely adopted as a common approach for networking connectivity and is typically implemented in various industrial scenarios. However, the rapid development of industry requires Ethernet to provide much faster transmission rates, such as the current capability of supporting up to 400 Gbps~\cite{miyamoto2012challenge}. Additionally, ongoing efforts aim to achieve speeds of 1 Tbps, known as Terabit Ethernet~\cite{tkach2011terabit,hui2008terabit}. 
%With the rapid industry development, Ethernet also requires some much faster transmission standards to transmit the data, e.g., current Ethernet can support up to 400 Gbps~\cite{miyamoto2012challenge}. However, that is not the end to speed up the transmission speed, some ongoing efforts  target reaching to 1 Tbps, which is known as Terabit Ethernet~\cite{TEthernet}.  Ethernet's best-effort service reduces network complexity and keeps protocol operation simple whilst maintaining protocol operation simple. 
%%%%%Despite widespread adoption in various industrial applications, Ethernet still lacks the deterministic QoS for e2e traffic. 
%TSN's overwhelming strength is its status as an open and standard technology that is not affiliated with any company or organization.

Notably, IEEE is not the only organization working on deterministic networks. IEEE has focused on TSN standardization, concentrating on the techniques of the TSN Task Group's physical layer (L1) and link layer (L2). Meanwhile, the Internet Engineering Task Force (IETF) formed the Deterministic Network (DetNet) Working Group~\cite{finn2017deterministic}, which focuses on network layer (L3) and higher-layer techniques. 
In the following sections, we will discuss TSN and its standardization efforts in meeting the requirements of industrial automation. 

\subsection{Time-Sensitive Networking (TSN)}\label{ssec:tsn}
One of the goals of TSN design is to guarantee the timing behaviors of critical traffic by temporarily isolating non-critical traffic. The IEEE 802.1 TSN Task Group (TG) focuses on this objective. 
TSN originates from the IEEE 802.1 AVB set of standards, which provide features such as low-latency traffic flow specifications, bandwidth reservations to guarantee a certain amount of traffic bandwidth across the network, and network management protocols to support synchronized traffic management. However, most of AVB's efforts focus on real-time efficiency within IEEE 802.1 networks, which is still insufficient to support the broad range of time-sensitive and mission-critical traffic flows in industrial automation~\cite{bello2019perspective}. The TSN TG standardizes these features to close the performance gap that remains in current systems. TSN offers several advantages to automation industries that have struggled for years with various incompatible proprietary communication protocols: 

$\bullet$ TSN ensures vendor-independent interoperability for all features of an industrial system. This provides users with more options to select devices for their schemes, avoiding lock-in by vendors and allowing system-wide connectivity.

$\bullet$ {\rev TSN addresses scalability issues since it is based on Ethernet, which is highly scalable in end stations and bridges. }
%Most existing TSN protocols are based on the Ethernet standard; it automatically scales with the Ethernet, meaning that the bandwidth or other performance criteria of the technology are not restricted. 
It is easy to add new nodes to the network and discover them through standard Ethernet protocols.

$\bullet$ {\rev TSN provides higher flexibility through its standardized technology, enabling the network structure to be flexibly extended without compatibility issues.}
%TSN can be used to improve the communication efficiency between machines and corporate systems. For example, it can establish communication among existing Ethernet and the TSN Ethernet without involving direct modifications to the existing non-critical communication infrastructure. 

{\rev As an open IEEE standard, TSN can not only ensure seamless communication between devices from different manufacturers, but also be integrated with other technologies in the higher layers of the OSI model, such as OPC Unified Architecture (OPC UA), another open and vendor-independent standard. These properties allow for greater interoperability, scalability, and flexibility in industrial automation systems.
}

\begin{figure}[tb]
  \centering
  \includegraphics[width=0.6\columnwidth]{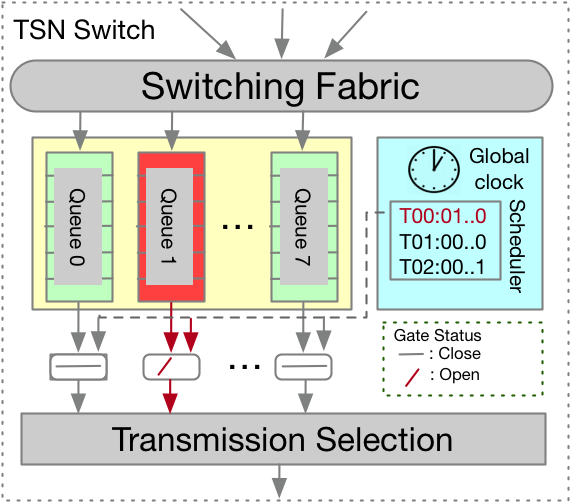}
  \caption{An illustration of a TSN switch. It consists of four key components: the switching fabric, the queues (each equipped with a gate), a global scheduler, and the transmission selection. The gate can only transmit in the ``open" state. }
  \label{Fig:TSNSwitchz}
  %\vspace{-0.15in}
\end{figure}

{\rev To realize the many features TSN provides, the design of a TSN switch plays a fundamental role in making traditional Ethernet have real-time characteristics. 
}
%Time synchronization and traffic scheduling are the key features provided by TSN. In industrial automation, these features can be addressed by the IEEE 802.1AS and IEEE 802.1Qbv standards, respectively. In general, all TSN-enabled devices, e.g., {\rev TSN bridges (or TSN switches)}\footnote{{\rev In this paper, we use TSN switch and TSN bridge interchangeably.}}, will be synchronized with a universal clock, and each switch has a timing table for all traffic types, and the switch schedules this traffic according to specifications (for example, traffic priority). 
{\rev A TSN switch is built on a gate driver mechanism and consists of multiple queues per port to buffer traffic with different priorities. The forwarded traffic is scheduled according to the control of each gate by carefully determining the time of its opening/closing. Such a mechanism guarantees that the communication delay is predictable and can be managed in a deterministic way. }
%will be held until its corresponding gate is open at a scheduled time slot. Once the gate is open, the queued transmission traffic will be released. 
%The timed release of packets guarantees that the network delay can be predicted and managed in a deterministic way. This behavior enables critical traffic and non-critical traffic convergence on the same network, where transmission of multiple priorities (including real-time and best-effort traffic) is via the same links. Different traffic classes can co-exist on the same network with no impact on higher criticality level traffic from traffic with lower priority~\cite{IntelWP}. 
Fig.~\ref{Fig:TSNSwitchz} shows an abstract of the TSN switch. It consists of four key components: the switching fabric to filter the traffic, the queues (each equipped with a gate) to buffer the traffic, a global scheduler, and the transmission selection. %The gate can only transmit in the ``open" state. 

{\rev Based on the gate driver switch architecture, TSN defines a collection of standards and amendments to meet the demands of industrial automation, especially the deterministic communication of critical traffic in the converged networks. At the highest level, by resource reservation and applying various queueing and shaping technologies, TSN achieves zero congestion loss for critical traffic, and this in turn, allows a guarantee on the e2e latency. TSN also provides ultra-reliability for critical traffic via frame replication as well as protection against bandwidth violation, malfunctioning, and malicious attacks~\cite{farkas2018time}. In addition, TSN supports frame preemption, which, on one hand, reduces the latency of critical traffic and on the other hand, improves the efficiency of bandwidth usage for non-critical messages. 
}

\begin{figure}[tb]
  \centering
  \includegraphics[width=\columnwidth]{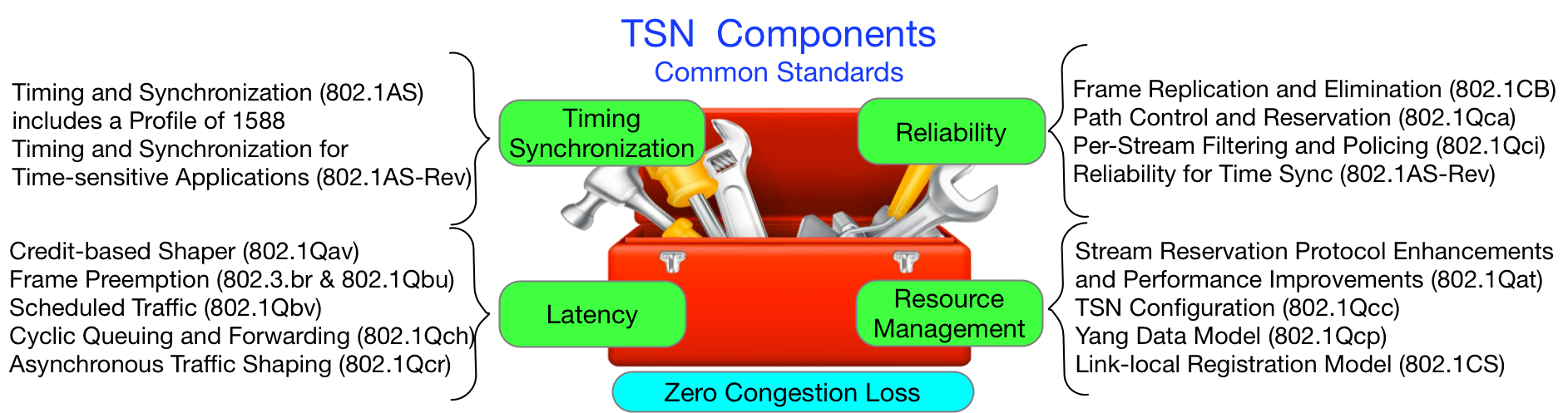}
  \caption{IEEE 802.1 TSN toolbox, consisting of four coarse sub-classes. {\rev Some draft standards (e.g., IEEE P802.1ASdm and IEEE P802.1Qdj) are still in progress and thus are not included in the discussion in Section~\ref{Sec:Chal}.}}
  \label{Fig:Toolset}
  %\vspace{-0.15in}
\end{figure}

The TSN standards provide a flexible toolbox from which a network designer can pick what is required for designing the targeted application. However, each protocol in this toolbox may not exist independently, and some competing approaches to configuring individual protocols are mutually exclusive and only support individual protocol feature sets.
As an overview, here we list some relevant TSN specifications for industrial automation~\cite{TSNClass}, as shown in Fig.~\ref{Fig:Toolset}. Their details will be provided in Section~\ref{Sec:Sta}.

$\bullet$ \textit{IEEE 802.1AS(-Rev)}
``Timing and Synchronization for Time-Sensitive Applications" and its revision (IEEE 802.1AS-Rev) are key TSN standards for achieving time synchronization among network components, essential for deterministic transmission in most TSN implementations. IEEE 802.1AS~\cite{5741898} includes several versions that utilize the IEEE 1588 Precision Time Protocol (PTP)~\cite{4579760} as the primary profile for synchronization~\cite{teener2008overview}. This protocol serves as the foundational standard for TSN traffic scheduling, relying on precise timing and synchronization among devices. The amended version, IEEE P802.1AS-Rev (e.g., P802.1AS-Rev/D8.3 drafted in Oct. 2019~\cite{8979571}), includes enhancements such as support for fault tolerance and scenarios with multiple active synchronization masters.

%$\bullet$ \textit{IEEE 802.1AS(-Rev)} 
%``Timing and Synchronization for Time-Sensitive Applications" and its upcoming revision are the set of TSN standards for time synchronization among the network components. Timing and synchronization are essential components to achieve the deterministic communication. IEEE 802.1AS uses the IEEE 1588 Precision Time Protocol (PTP) as primary profile to enable compatibility between the distinct TSN systems for synchronization~\cite{teener2008overview}.  IEEE 802.1AS also sets the foundation for traffic scheduling through each network device involved. IEEE 802.1AS-Rev, an amendament version of IEEE 802.1AS, is used to support the fault tolerance in time synchronization, as well as the case of multiple active synchronization masters.

$\bullet$ \textit{IEEE 802.1Qbv} 
``Enhancements for Scheduled Traffic", also known as the ``time-aware shaper (TAS)"~\cite{al2017modeling}, introduces the concept of a time-triggered (TT) switch with multiple virtual queues. With the help of a common shared global scheduler, established by centralized network configuration (CNC) (for the details please refer to IEEE 802.1Qat~\cite{5594972} and IEEE 802.1Qcc~\cite{8514112}), IEEE 802.1Qbv~\cite{8613095} controls the open or closed status of the gates at the egress of a switch to manage the flow of queued traffic. A typical TSN switch is a priority-based switch where different priority traffic types go into different priority queues, each equipped with a gate. A queue can only transmit traffic when its gate is open. By following a strict schedule, the delay is deterministic at each switch, ensuring that the e2e latency is guaranteed in a TSN-enabled network. %In general, a TAS shaper can protect critical traffic from interference from other network traffic. However, it does not optimize the use of bandwidth or minimize the latency of communication~\cite{TSNQbu}. 

%$\bullet$ \textit{IEEE 802.1Qbv} 
%``Enhancements for Scheduled Traffic" introduces  the time-triggered (TT) switch over between virtual egress queues of a switch (also known as the ``time-aware shaper"~\cite{al2017modeling}).  Depending on the common shared global time provided by IEEE 802.1AS standard, a schedule can be established and distributed among the participated network devices.  IEEE 802.2Qbv defines mechanisms at the egress of a TSN switch to control the flow of the queued traffic through gates. During the scheduled time windows, the transmission of messages from these queues is executed. However, other lower priority queues will be blocked from these transmission, which eliminates the chance that higher priority scheduled traffic is being influenced by the non-scheduled traffics. By doing so, the delay is deterministic at each switch and the e2e message latency can be guaranteed in a TSN-enabled network.

$\bullet$ \textit{IEEE 802.1Qav} ``Forwarding and Queuing Enhancements for Time-Sensitive Streams", known as the ``credit-based shaper (CBS)"~\cite{mohammadpour2018latency}, is designed to limit the bandwidth for multiple stream transmissions. By collaborating with the stream reservation protocol (SRP)~\cite{bello2014novel}, the CBS shaper can manage the buffer size at the receiving port, providing bounded latency per stream type. Additionally, IEEE 802.1Qav can restrict the transmission of audio/video frames to protect best-effort traffic. %ensuring that critical streams have the necessary bandwidth while minimizing interference with non-critical traffic. 

%\bullet$ \textit{IEEE 802.1Qav} ``Forwarding and Queuing Enhancements for Time-Sensitive Streams" introduces the feature of bandwidth limitations for multiple stream types in a network (also known as the ``credit-based shaper (CBS)"~\cite{mohammadpour2018latency}). By combining the stream reservation protocol (SRP)~\cite{bello2014novel} to reserve the time slots and CBS shaper together, it is possible to limit the requirement on the size of buffer at the receiving port, and this further can provide a bounded latency per stream type.  The CBS spaces the A / V frames in order to decrease bursting and bunching, thus protecting best-effort traffic as the maximum AVB burst is confined. 

$\bullet$ \textit{IEEE 802.1CB} 
``Frame Replication and Reliability Elimination" provides a mechanism for duplicating streams to enhance reliability, e.g., transmitting a stream over multiple available paths and re-merging the duplicates at the destination port. The redundancy management in IEEE 802.1CB~\cite{8091139} follows a scheme similar to those established by the High-availability Seamless Redundancy protocol and Parallel Redundancy Protocol~\cite{8073824}. Utilizing IEEE 802.1Qca, also known as ``Path Control and Reservation", the redundancy management in IEEE 802.1CB can set up and manage designated disjoint paths, thereby maintaining full control over the duplicated streams.

%$\bullet$ \textit{IEEE 802.1CB} ``Frame Replication and Reliability Elimination" offers a means of duplicating streams, e.g., sending them over multiple paths and re-merging duplicates back into a single stream. Redundancy management in IEEE 802.1CB follows the similar approaches established from High-availability Seamless Redundancy protocol and Parallel Redundancy protocol~\cite{8073824}. For example, to increase the success of the message transmission, the transmitted message are duplicated and transmitted in parallel over multiple disjoint paths. The standard IEEE 802.1Qca (called ``Path Control and Reservation") specifies the mechanism to set up and manage the designated disjoint paths. Combining these received and duplicated messages into a single stream on the receiver side is the task of the redundancy management system.

$\bullet$ \textit{IEEE 802.1Qcc} 
``Stream Reservation Protocol Enhancements and Performance Improvements" offers various models for reserving streams on a TSN-enabled network. It supports three resource management models: a fully distributed model, a centralized network/distributed user model, and a fully centralized model. This protocol enables deterministic stream reservation on each intermediate bridge, thereby guaranteeing e2e latency. %\sout{(3.5) In general, for automation industries, the fully centralized model is the most suitable. }

%$\bullet$ \textit{IEEE 802.1Qcc} ``Stream Reservation Protocol Enhancements and Performance Improvements" provides three models on configuring the parameters of other TSN standards.  It provides several key features to reserve the streams, e.g., enabling  multiple simultaneous transmitted streams, supporting the configurable stream reservation and the deterministic stream reservation convergence, and adding the support on routing and reservations for User-Network Interface (UNI). Besides, offline and/or online configurations are also supported for the TSN network schedule. The two configuration models currently under examination are fully centralized and distributed models~\cite{gutierrez2017self}.

$\bullet$ \textit{IEEE 802.1Qbu} ``Frame Preemption" (together with IEEE 802.3br ``Specification and Management Parameters for Interspersing Express Traffic"~\cite{7592835}) provides a mechanism allowing higher priority frames to interrupt lower priority frames. This ensures that critical traffic is protected from interference by non-critical traffic. {\rev Although the TAS shaper in IEEE 802.1 Qbv can mitigate transmission jitter by blocking lower priority queues before the transmission begins, the preemption capability defined in IEEE 802.1 Qbu~\cite{7553415} is essential for further enhancing the real-time performance of critical traffic. }%However, it does not optimize the use of bandwidth or minimize the latency of communication~\cite{TSNQbu}. 
%For example, a low priority frame is a larger frame, which takes longer to finish its transmission. With IEEE 802.1Qbu, the high priority frame can intercept these larger frames.  Once the high priority frame is finished, the low priority frame can resume. 

%$\bullet$ \textit{IEEE 802.1Qbu} (together with IEEE 802.3br) ``Frame Preemption" (together with'' Specification and Management Parameters for Interspersing Express Traffic"), e.g., the switch-related part and the endpoint-related part of the frame preemption, provides a mechanism to allow higher priority frames to interrupt the lower priority frames (which is currently being sent to the same egress port) at every multiple of 64 Bytes (which are the allowed minimum size of an Ethernet frame). While this mechanism protects higher-priority (or critical) messages from interference of other network traffics, it does not aim to optimize the use of bandwidth or minimize the latency of communication. IEEE 802.1Qbuu can interrupt the transmission of the standard Ethernet or larger frames to allow the higher priority frames to be transmitted, and it then can resume, instead of discarding, the transmitted pieces of the previously interrupted message.

{\rev In summary, the IEEE TSN TG focuses on enhancing the reliability and real-time capabilities of the Ethernet standard in industrial automation through a comprehensive set of standards. This includes IEEE 802.1AS for time synchronization, IEEE 802.1Qbv/802.1Qbu/802.3br for traffic shaping and scheduling, IEEE 802.1CB for reliability, and IEEE 802.1Qcc for centralized resource management. In the following, we will delve deeper into how TSN achieves high-precision time synchronization, bounded latency, reliability, and resource management through these standards. 
}

%% file: sec/TSN.tex
\section{TSN Standardization}
\label{Sec:Sta}
%This section provides the details of the standardization efforts of IEEE 802.1 TSN TG. TSN targets at enhancing and improving the data-link layer performance of the traditional Ethernet to meet the requirements of industrial automation applications. 
%As described in the previous section, TSN offers a flexible toolset, allowing network designers to select the tools needed for specific applications. 
%Additionally, TSN can be integrated into Industrial Automation and Control System (IACS) systems to provide more robust services over existing Ethernet infrastructure~\cite{xu2000industrial}. 
%In this section, we categorize the TSN standardization efforts into different classes within industrial automation. From a system design perspective, a typical TSN flow involves multicast or unicast Ethernet links, comprising a source end station (also called a talker or sender) and a sink end station (also called a listener or receiver), both connected over a TSN-enabled network. The primary goal of TSN is to define protocols that connect these end devices via bridges, ensuring they function correctly according to the specified standards. 

We can broadly classify the TSN standardization efforts into four major sets, as shown in Fig.~\ref{Fig:Toolset}, while the classifications are not disjoint, as some standards contribute to multiple aspects. %We characterize a TSN flow by the QoS properties defined for the traffic class to which the flow belongs, e.g., bandwidth and latency. 
The four main pillars on which TSN is built are: 1) time synchronization (e.g., with IEEE 802.1AS, which includes a profile of IEEE 1588), 2) guaranteed e2e latency (e.g., scheduled traffic defined in IEEE 802.1Qbv), 3) reliability (e.g., with redundancy provided by IEEE 802.1CB and per-stream filtering by IEEE 802.1Qci) and 4) resource management (e.g., configuration in IEEE 802.1Qcc and YANG models in IEEE 802.1Qcp).  We will detail each aspect below,  
{\rev and explain the advantages of TSN over the existing industrial solutions at the end of this section. }

\subsection{Time Synchronization}\label{ssec:sync}
Time synchronization is crucial for most applications targeted by the IEEE 802.1Q standards. Many TSN standards depend on network-wide precise time synchronization, with varying requirements when transitioning from AVB streaming to time-sensitive and safety-critical control applications. In a typical TSN network, a common time reference is shared by all TSN entities and used to schedule data and control signaling. Time synchronization in TSN is defined primarily by two key standards: IEEE 802.1AS and IEEE 802.1AS-Rev. %However, there are some exceptions, such as IEEE 802.1Qcr Asynchronous Traffic Shaping (ATS), which can operate in an asynchronous setting.

The IEEE 802.1AS standard utilizes and optimizes the IEEE 1588-2008 (1588v2) protocol, which includes the Generic Precision Time Protocol (gPTP) to synchronize clocks across the network~\cite{stanton2018distributing}. It is also one of the three IEEE 802.1 AVB standards, targeting network audio/video applications. gPTP achieves clock synchronization between network devices by exchanging predefined messages across the communication medium~\cite{levesque2016survey}. %It includes the transport of synchronized time over bridged / virtual bridged local area networks, the selection of the timing source, and the notification of timing requirements, such as phase and frequency discontinuities. 

%\begin{figure}[tb]
%  \centering
%  \includegraphics[width=0.4\columnwidth]{Figs/AS.png}
%  \caption{The Master-Slave Hierarchy established by the BMCA algorithm~\cite{eidson2002ieee}. TIA represents International Atomic Time, which provides the timing information to GrandMaster (GM).}
%  \label{Fig:AS}
%  \vspace{-0.15in}
%\end{figure}

A typical gPTP employs a messaging mechanism between the Clock Master (CM), also known as the GrandMaster (GM), and Clock Slaves (CS) to create a time-aware network. This network uses peer-path delay to calculate timing information such as link latency (between bridges) and residence time (within bridges). Link latency consists of the time spent on the link (e.g., the single-hop propagation delay between two adjacent switches), and residence time includes the time spent within the switch (e.g., processing time, queuing time, and transmission time). 
The GM clock serves as the reference time at the root of the time-aware network hierarchy and is selected by the Best Master Clock Algorithm (BMCA)~\cite{yu2009best}, which automatically designates the grandmaster device. The BMCA dynamically configures the synchronization hierarchy, known as the synchronization spanning tree. This spanning tree is a minimum spanning tree (MST) constructed by assigning each port to one of three states: master, slave, or passive, with an additional disabled state for ports not in use. %For example, some work~\cite{van2003lightweight} constructs a lightweight time synchronization spanning tree. 
In the gPTP protocol, devices are categorized as gPTP-enabled or non-gPTP-enabled, distributed, and interconnected within the network. gPTP-enabled devices, including time-aware bridges and time-aware end stations, contribute to time synchronization, while non-gPTP-enabled devices do not need to provide these features.
%To achieve time synchronization, the gPTP takes a two-phase process: (a) setting up a master-slave hierarchy, and (2) applying clock synchronization algorithm. gPTP uses the BMCA to establish a master-slave hierarchy consisting of two separate algorithms: data set comparison and state decision. Once the synchronization hierarchy has been established, gPTP defines how the time formed by the root (e.g., the grandmaster) can be distributed to the other devices in the network. Each gPTP device operates a gPTP engine, such as a gPTP state machine, and uses multiple gPTP UDP/IP multicast and unicast messages to set the appropriate hierarchy and synchronize time correctly. In general, the non-time-aware bridges are not used to calculate the BMCA clock, and they will not be involved to generate the spanning tree since these bridges cannot provide the timing synchronization information. 

IEEE 802.1AS-Rev introduces new capabilities required for time-sensitive applications in several ways. First, GMs and synchronization trees can be redundantly configured to enhance fault tolerance, allowing synchronization trees to be explicitly configured without using the BMCA algorithm. Additionally, IEEE 802.1AS-Rev supports redundant communication by enabling multiple time domains for gPTP. Each gPTP domain operates as a separate instance, allowing network devices to execute multiple instances of gPTP simultaneously. This enhances redundancy by permitting multiple grandmaster clocks and synchronization spanning trees, facilitating seamless low-latency transfer. % and quick recovery if one synchronization process is failed.

\subsection{Bounded Latency}\label{ssec:tsnlatency}
One primary characteristic of TSN standards is the guaranteed delivery of messages with stringent timing constraints, i.e., bounded e2e latency. %The e2e latency of flows is typically calculated based on traffic priorities and arrival patterns. %, which makes compositional system design and time behavior isolation very hard. 
The IEEE 802.1Qav standard has been created as part of the AVB performance enhancement suite {\rev to specify data transmission protocols and bridge mechanisms.} In addition to IEEE 802.1Qav, TSN provides more standards, such as IEEE 802.1Qbv and IEEE 802.1Qbu, to further enhance real-time performance. In this section, we discuss several standards in TSN towards bounded latency. 
%which specifies protocols and mechanisms. % that ensure real-time transmissions even without the worldwide end station and the  bridge synchronization.  
%In addition to IEEE 802.1Qav, TSN provides more standards, e.g., IEEE 802.1Qbv and IEEE 802.1Qbu, to further enhance the real-time performance. %IEEE 802.1Qbv defines schedule-driven communication, e.g., to leverage synchronized time in network transmission and forwarding decisions. While 802.1Qbu specifies a mechanism for preemption that enables time-critical or high-priority messages to interrupt ongoing non-time-critical or low-priority messages. 
%TSN flows are typically based on the schedule-driven communication which is often referred to as time-triggered communication. %The key principle of TT communication is straightforward: system designers generate a communication schedule that instructs the end stations to send specified frames to the network, this communication schedule is distributed as part of the configuration of the end stations and bridges, and a scheduling function at the end station executes the communication schedule. 

\subsubsection{IEEE 802.1Qav Forwarding and Queuing of Time-Sensitive Streams}

{\rev IEEE 802.1Qav specifies the enhancements for the transmission selection algorithms of Ethernet switches and defines the credit-based shaper (CBS) to ensure bounded latency for time-sensitive traffic by regulating the transmission rate. CBS is a traffic shaping mechanism that regulates bandwidth allocation for high-priority shaped queues to reduce delays in medium- and low-priority unshaped queues, thereby enhancing fairness. 
}
%SP is used as the default algorithm for selecting legacy Ethernet frames for transmission, which should be supported by all switches, and these frames are typically selected by their priority values. Together with the Stream Reservation Protocol (SRP), CBS can be used to limit the amount of buffering required in an IEEE 802.1Qav receiving station. The CBS protocol separates the AV traffic from other traffic to reduce the chance of bursting and bunching, which potentially protects the best-effort traffic as the maximum audio/video traffic burst is confined. %{\gw And this will give best-effort traffic a fair chance to transmit these traffic. }
%Also, by limiting the back-to-back AVB stream bursts, it can prevent congestion in a downstream bridge. 

{\rev In CBS, each output queue is associated with a credit counter. The credit counter accumulates credits when the queue waits to transmit frames and consumes credits when frames are transmitted. A frame can only be transmitted if the credit of its queue is non-negative and no other frames are being transmitted at the same time. If no frames are waiting for transmission, the credit of the queue is reset to zero. The queue credit decreases and increases at a constant rate, specified by the configurable parameters sendSlope and idleSlope, respectively. 
} %This guarantees no conflicting frames existing~\cite{lim2012performance}. 
%If a frame is dequeued and transmitted, it will decrease its queue credits at the rate of \textit{sendSlope}. In the other case, e.g., if there is at least one frame in the queue, the credits are increased at the rate of \textit{idleSlope}. 
%If a waiting frame is dequeued and scheduled to transmission, it decreases its queue credits at a constant rate, specified by the \textit{sendSlope}. Otherwise (e.g., if there is at least one frame in the queue), the credits of waiting frames are increased at a constant rate, specified by the \textit{idleSlope}. 
idleSlope measures the actual bandwidth reserved for a specific queue within a bridge, while sendSlope reflects the port transmitting rate~\cite{lim2012performance}. By setting the parameters hiCredit and loCredit, the CBS shaper can limit the maximum and minimum values at which credits can be accumulated. 
%By setting the parameters \textit{hiCredit} and \textit{loCredit}, the CBS shaper can limit the maximum and minimum values at which credits can be accumulated. %This potentially increases the algorithm's robustness.
In general, the CBS shaper does not negatively impact the utilization rate of network links. If CBS delays the transmission of high-priority frames, lower-priority frames can be transmitted if they are queued.  When CBS is used with resource reservation and admission control, it ensures that network utilization does not drop below the requested rates. %Although it may limit the available bandwidth for real-time traffic, this is intended to prevent starvation of lower-priority traffic, rather than offering guaranteed latency.

%The CBS shaper separates a queue into two traffic classes, e.g., class A with a tight delay bound and class B with a loose delay bound. 
%In a CBS shaper, the transmission is only permitted if the credits given in bits are greater or equal to zero and no other frames are transmitted simultaneously (e.g., no conflicting frames). 

%\begin{figure}
%  \centering
%  \includegraphics[width=0.5\columnwidth]{Figs/CBS2.png}
%  \caption{An example of the Credit-based Shaper Operations. When a frame is sending, its credit decreases, while when it becomes idle, its credit increases~\cite{queck2012analysis}. }
%  \label{Fig:CBS}
%  \vspace{-0.15in}
%\end{figure}

%Fig.~\ref{Fig:CBS} shows an example describing the operation of the CBS shaper. Initially, a single AVB frame, Frame A, arrives and is queued because of the conflicting frames, and its credits are increased at a rate of \textit{idleSlope}. Once the previous transmission is finished, Frame A is selected for transmission and its credits will be decreased at a rate of \textit{sendSlope}. %If there are no frames for transmission, the credits are set to zero and the algorithm should wait for the next transmission. 
%Another frame, Frame B, comes in the second slot and is immediately transmitted due to its credit values ($>=0)$ and absense of conflicting frames. Its credits are decreased at a rate of \textit{sendSlope} and increased again at the rate of \textit{idleSlope} after the transmission completes. 
For bandwidth-intensive applications, the CBS protocol can establish an upper bound for each traffic class, ensuring that no traffic class exceeds the pre-configured threshold on reserved bandwidth, typically less than 75\% of the maximum bandwidth. Along with SRP, the CBS shaper aims to limit delays to less than 250 $\mu$s per bridge and the worst-case latency to up to 2 $ms$ for class A, and up to 50 $ms$ for class B in a simple network setup~\cite{lim2012performance}.
However, these delay scales may still be too high for industrial control applications. This has motivated the TSN TG to introduce other standards, such as IEEE 802.1Qbv, IEEE 802.1Qch, and IEEE 802.1Qcr, to meet the stringent timing requirements of industrial applications.

\begin{figure}
  \centering
  \includegraphics[width=0.9\columnwidth]{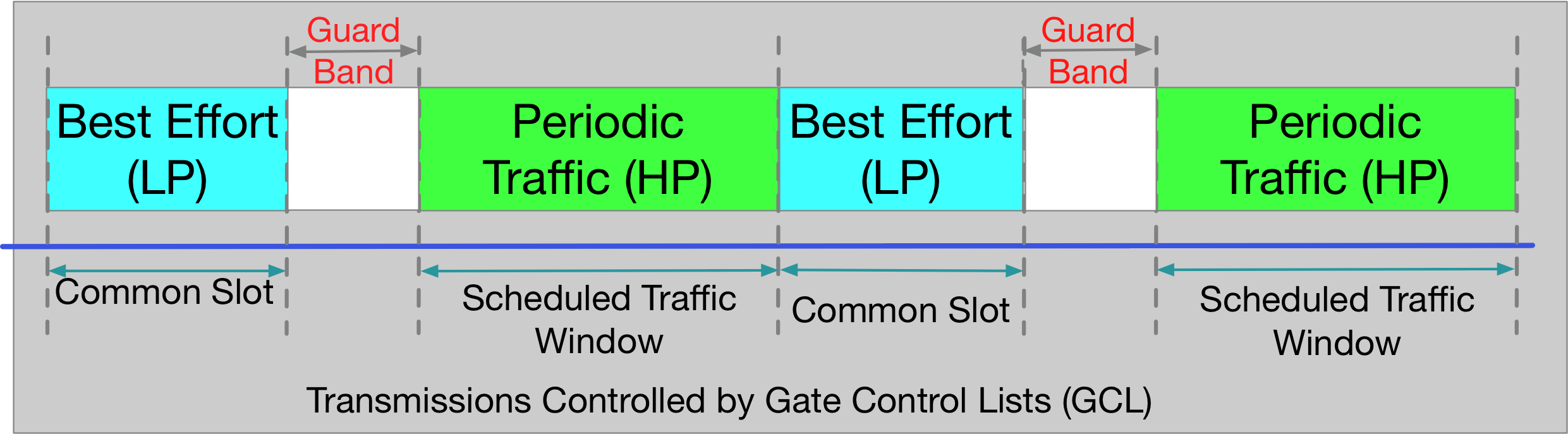}
  \caption{Overview of IEEE 802.1Qbv Time-Aware Shaper (TAS): the scheduled traffic will be sent over TDMA-like synchronized slots. HP traffic have guaranteed reserved resources across the network, while LP traffic are best-effort low-priority traffic. }
  \label{Fig:TAS}
  %\vspace{-0.15in}
\end{figure}

\subsubsection{IEEE 802.1Qbv Enhancements to Traffic Scheduling (Time-Aware Shaper (TAS))}\label{sssec:tas}
%{\gw (3.6) While 802.1Qav applies coarse-grained per-priority transmission selection, and the CBS shaper is applied per-queue. }
%Unlike IEEE 802.1Qav, 802.1Qbv does not directly schedule the frame transmission. 
%IEEE 802.1Qav applies coarse-grained per-priority transmission selection by default, 802.1Qbv schedules the activation and deactivation of the queues. 
IEEE 802.1Qbv introduces the concept of a gate per queue as a means of realizing the activation and deactivation of a queue. The gate is in the open or closed state to activate or deactivate the queue, respectively. The transmission selection process at the port of a bridge may only select frames from queues that are activated, i.e., having their gates in the open state. The combination of the TAS shaper with the frame preemption scheme is desirable for supporting deterministic traffic in industrial applications, such as mission-critical applications with sub-microsecond latency.

In TAS, critical traffic is scheduled in protected traffic windows with allocated time slots to transmit the traffic, similar to the TDMA paradigm. Each window can have an allotted transmission time for high-priority traffic, as illustrated in Fig.~\ref{Fig:TAS}. To prevent potential interference among the assigned time slots, the traffic windows should be protected with minimal time requirements, called the guard band. The guard bands enforce time intervals after best-effort traffic during which all gates are closed, ensuring neither best-effort traffic nor periodic traffic can be sent during these intervals. These guard bands are required to prevent large best-effort frames from interfering with periodic traffic. 

The TAS shaper requires that all traffic windows be well synchronized and scheduled among all the time-aware bridges. {\rev The communication schedule in IEEE 802.1Qbv is realized by the scheduled gate mechanism, which controls the opening and closing of queues using a pre-determined gate control list (GCL). Each GCL includes a limited number of entries, with each entry providing the status of associated queues over a particular duration. The GCL repeats itself periodically, and this period is called the cycle time. The network-wide schedule is generated by centralized network configuration (CNC) and deployed on individual bridges. Although the IEEE 802.1Qbv standard defines the scheduling mechanism of TAS, its configuration, i.e., what to put in the GCL and how to assign queues for individual traffic at each hop, lacks a clear-cut best practice~\cite{xue2023real}. This has resulted in significant efforts from both researchers and practitioners to study the TAS-based scheduling problems in various industrial applications. More discussion regarding TAS scheduling is provided in Section~\ref{ssec:scheduling}.
}

\subsubsection{IEEE 802.3br and 802.1Qbu Interspersing Express Traffic and Frame Preemption}
To address the inverted priority problem, i.e., ongoing transmission of a low-priority frame prevents the transmission of high-priority frames, the IEEE 802.1 TG along with the IEEE 802.3 TG defined the frame preemption protocol in IEEE 802.1Qbu and IEEE 802.3br. 
{\rev These technologies work together to effectively manage traffic using changes to both the MAC scheme, as controlled by IEEE 802.3, and management mechanisms, as supervised by IEEE 802.1. The frame preemption capability can be combined with any traffic management algorithms defined in IEEE 802.1Q, such as the TAS shaper and CBS shaper, to enhance determinism and real-time performance for critical traffic. }

%IEEE 802.1Qbu defines the procedures for bridges and end stations to hold or suspend the transmission of a frame in order to allow the transmission of one or multiple more urgent frames, as well as procedures to release or resume the transmission of the less critical frames once the urgent frames have been transmitted. 

The IEEE 802.1Qbu standard allows urgent and time-critical data frames to be split into smaller fragments and preempt the non-critical frames on the same physical link, even if they are in transition. This frame preemption scheme divides an egress port into two distinct interfaces based on the MAC layer: preemptable MAC (pMAC) and express MAC (eMAC)~\cite{park2019design}. The pMAC targets preemptable frames, while the eMAC targets preemptive frames. An incoming frame is mapped to only one egress interface according to the frame preemption status table, with the default option being the eMAC. In general, a pMAC frame can be preempted by an eMAC frame even when the pMAC frame is in transition. Only after the eMAC frame completes its transmission can the pMAC frame resume its transmission.  

%http://www.ieee802.org/3/br/Comments%20received%20on%20IEEE%20P802.3br%20drafts/IEEE_P802d3br_figures_DL.pdf
The IEEE 802.3br standard introduces an optional sublayer called the MAC Merge sublayer, which attaches an eMAC and a pMAC to the PHY layer through a reconciliation sublayer~\cite{zhou2017analysis}. The PHY layer remains unaware of the preemption, while the MAC Merge sublayer and its MACs support frame preemption as defined in IEEE 802.1Qbu. The MAC Merge sublayer provides two approaches to manage the transmission of preemptable traffic alongside express traffic. One approach interrupts (preempts) the preemptable traffic currently being transmitted, while the other prevents preemptable traffic from being transmitted in the first place. However, frame preemption typically induces some overhead due to the content switching between two distinct frames. Unlike the guard bands defined in IEEE 802.1Qbv for scheduled traffic (as shown in Fig.~\ref{Fig:TAS}), frame preemption does not require guard bands. When used together with scheduled traffic, frame preemption can reduce the required size of the guard band from the maximum transmission unit (MTU) size of 1522 bytes to a smaller size of 127 bytes. %The preemption capability is only enabled for transmission after verifying that the other end of the link also supports preemption. %The assessment is made during a verification phase, which follows a negotiation based on the link layer discovery protocol and the exchange of additional Ethernet parameters. In particular, each traffic class queue supported by each port of a bridge or end station is assigned one frame preemption status value (either express or preemptable) via the frame preemption status table. All priorities that are mapped to the same traffic class have the same preemption status value. %Preemption values can be changed by management. 

%Any of the available transmission selection algorithms defined in IEEE 802.1Q, such as CBS shaper, can be used with preemption enabled. The IEEE 802.1Qbu standard specifies the behaviors of the CBS shaper algorithm when used in combination with frame preemption. 

\subsubsection{IEEE 802.1Qch Cyclic Queuing and Forwarding (CQF)}

%\begin{figure}[tb]
%  \centering 
%  \subfigure[An example of CQF scheme]{ 
%    \label{Fig:CQF} 
%    \includegraphics[width=2.2in]{Figs/CQF.png} 
%  } 
%  \subfigure[An example of ATS bridge]{ 
%    \label{Fig:ATS} 
%    \includegraphics[width=3in]{Figs/ATS.png} 
%  } 
%  \vspace{-0.1in}
%  \caption{(a). An example of message (V) transmission CQF between two nodes at two consecutive cycles~\cite{qiang2019large}. (b) An example of ATS bridge: ATS shaper determines at the ingress the type of traffic. ST represents the highly-priority scheduled traffic, and BE represents the low-priority best-effort traffic~\cite{nasrallah2019performance}.} 
%  \vspace{-0.2in}
%\end{figure}

%IEEE 802.1Qav has some shortcomings, such as increased delay in pathological topologies, and both worst-case delays and buffer requirements in switches being topology-dependent. 
The IEEE 802.1Qch standard introduces the CQF mechanism, also known as the Peristaltic Shaper (PS)~\cite{thiele2015formal}, which synchronizes enqueue and dequeue operations to reduce e2e latency by minimizing the time data resides within Ethernet switches. PS achieves this by dividing the timeline into odd and even phases of equal widths. 
The synchronized procedures are typically performed cyclically at the local bridges, regardless of the network topology.
%For detailed timing analysis on PS, interested readers can refer to the following works~\cite{thangamuthu2015analysis,thiele2015formal}. 

\begin{figure}[tb]
  \centering
  \includegraphics[width=0.7\columnwidth]{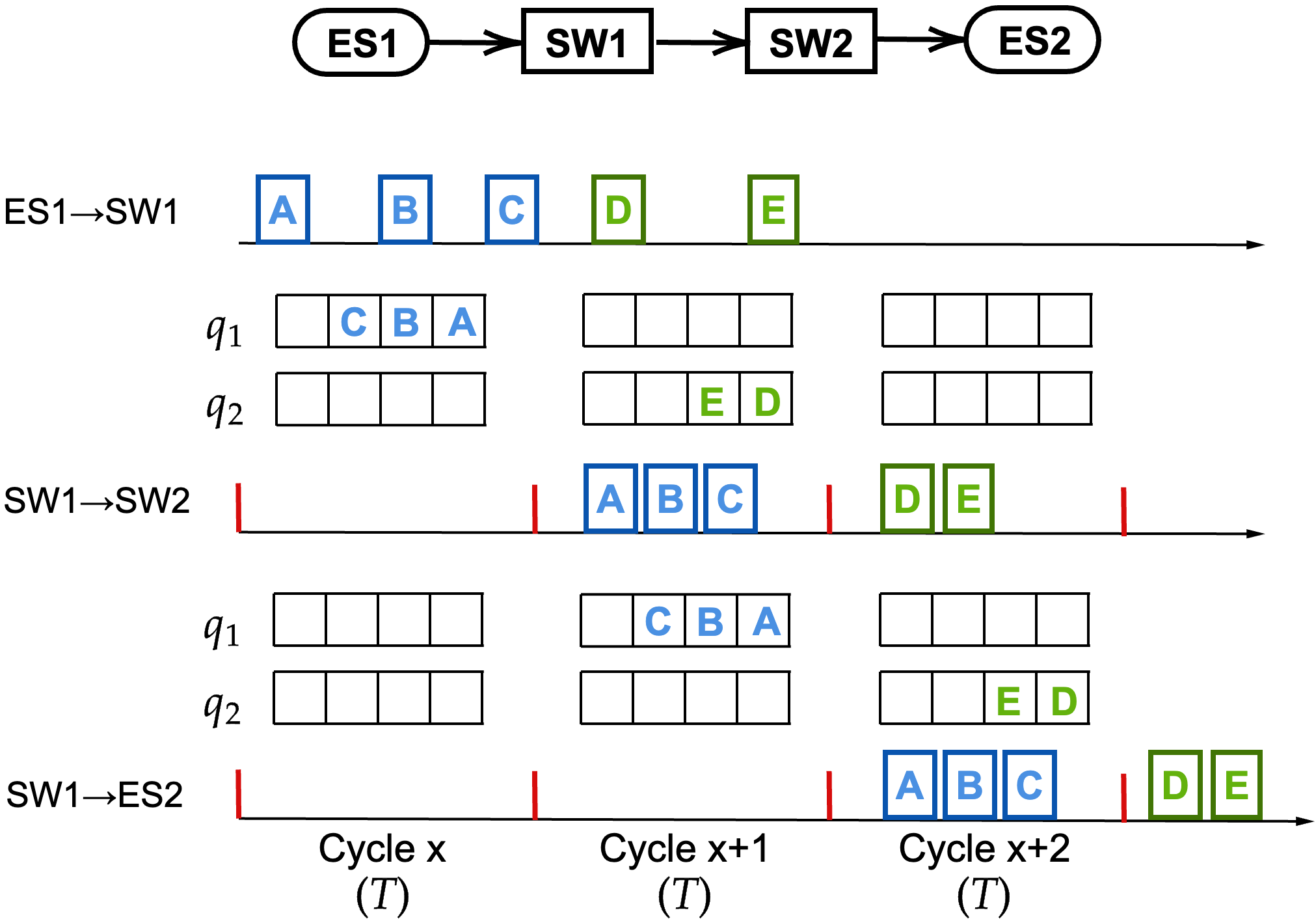}
  \caption{{\rev An example of CQF operation on a chain topology with two switches. Time is divided into cycles with the length of $T$. Frames received by a switch in cycle $x$ will be sent out in the next cycle $x+1$.}}
  \label{Fig:CQF}
  %\vspace{-0.15in}
\end{figure}

{\rev CQF is an efficient forwarding scheme proposed to simplify the design of a TSN switch, and it can deliver predictable and deterministic e2e latency~\cite{qiang2019large}. 
It is designed for limited-scale networks with time synchronization. Among the eight queues of a port of each switch, CQF reserves at least 2 queues performing enqueue and dequeue operations in a cyclic manner. Fig.~\ref{Fig:CQF} shows an example of CQF operation on a chain topology with two switches SW1 and SW2. Time is divided into equal cycles with the length of $T$ which is delimited by the red vertical lines. During the first interval (i.e., cycle $x$), frames $A$, $B$, and $C$ are sent out by end station ES1 and arrive at SW1, enqueueing them in $q_1$. In the following interval (i.e., cycle $x+1$), these frames are dequeued and forwarded to SW2, stored in $q_1$. Meanwhile, another two frames $D$ and $E$ arrive at SW1, enqueued in another queue $q_2$. The operation repeats in each cycle. CQF can provide a deterministic e2e latency guarantee since it follows two principles. 1) The sending cycle of a frame on a switch and the receiving cycle on the subsequent switch are the same. 2) Any frame received by a switch on cycle $x$ must be sent out on the next cycle $x+1$. Thus, the e2e latency of a frame is determined by the routing path length and cycle size $T$. }
%According to IETF standard~\cite{qiang2019large}, in CQF, ``each port needs to maintain 2 (or 3) queues, one receiving queue is used to buffer newly received packets, one sending queue is used to store the packets that are going to be sent out, one more queue may be needed to avoid output starvation". Fig.~\ref{Fig:CQF} shows an example of the CQF scheme, in which node A is the upstream node of node B. Under CQF, packets sent from A at the cycle $x$ will be received by B at the same cycle, then further be sent to the downstream node by B at cycle $x+1$. 
%Assuming that time-sensitive traffic in a synchronized setting, e.g., defined by IEEE 802.1AS or IEEE 802.1AS-Rev standards, is scheduled with a worst-case deterministic delay between the sender and receiver. Typically, a frame's transmission latency is defined only by the time cycle and the number of hops in its path. Thus, the transmission latency of a frame is completely independent of the topology and other non-TSN traffic parameters.

The frame preemption scheme can also work together with the CQF mechanism to improve transmission performance. This combination can potentially shorten the cycle time of frame transmission, as the size of a frame fragment is smaller than that of a full frame. To make CQF work properly, all frame fragments must be received within the scheduled time cycle. Accordingly, to guarantee bounded and deterministic latency, it is crucial to carefully design the cycle length along the routing path. 
{\rev Due to its simplicity, CQF can be easily supported by extending a standard Ethernet switch with statically configured queues, and several works (e.g.,~\cite{guidolin2023first,yan2020injection,wang2022joint}) have been proposed to study the configuration of CQF to optimize its deployment in practical TSN networks.}

\subsubsection{IEEE 802.1Qcr Asynchronous Traffic Shaping (ATS)}

%Although the combination of IEEE 802.1 CQF and TAS shaper can reduce the latency for the critical traffic (e.g., with high priority), due to the requirement of network-wide coordination, it is not efficient to utilize network bandwidth.  
%Generally speaking, two main approaches are undertaken when designing real-time Ethernet based on packet-switched networks: 1) synchronous time-triggered based medium access control (e.g., TAS), and 2) asynchronous event-triggered approach (e.g., ATS)~\cite{nasrallah2019performance}. 
{\rev The TAS shaper can provide deterministic real-time communication in a TSN network but requires high-precision network-wide time synchronization. However, industrial networks may suffer from timing misalignment, such as drift or skew in timing signal frames, lost timing frames, and inaccuracy, which can cause asynchrony. This issue worsens with the increasing scale of the network~\cite{zhou2019insight}.} To address this, IEEE 802.1Qcr aims to smooth out traffic patterns by reshaping TSN streams per hop and prioritizing urgent traffic over non-deterministic traffic. The ATS shaper works asynchronously, not requiring synchronization on traffic transmission, and relies heavily on an Urgency Based Scheduler (UBS). The UBS prioritizes urgent traffic by queuing and reshaping each individual frame at each hop. Asynchronicity is achieved through a Token Bucket Emulation (TBE) and an interleaved shaping algorithm to eliminate burstiness. The TBE controls traffic by the average transmission rate but allows a small portion of burst traffic to occur~\cite{specht2016urgency}.
%TAS works well under high precision time synchronization, however, TSN network may occur timing misalignment, e.g., drift or skew in timing signal frames, timing frames lost, and inaccuracy. All these events can cause the asynchronous in TSN timing synchronization. Also, with the network scale increase, this case is more serious~\cite{nasrallah2019performance}~\cite{zhou2019insight}.  
%IEEE 802.1Qcr aims to smooth out the traffic patterns by reshaping TSN streams per hop, and prioritizing the transmission of urgent traffic over non-deterministic traffic. In general, the ATS shaper works asynchronously, which does not require synchronization on traffic transmission.   
%ATS heavily relies on an Urgency Based Scheduler (UBS), which can prioritize the urgent traffic using queuing and reshaping of each individual frame at each hop. The asynchronicity is achieved by a Token Bucket Emulation (TBE) and an interleaved shaping algorithm to eliminate the burstiness. The TBE can control traffic by average transmission rate, but allows a small portion of burst occur (details refer to~\cite{specht2016urgency}). 
%This approach designed for cycle synchronization or time-triggered operations without relying on the network topology, and 
One aim of the ATS design is to provide deterministic and relatively low transmission delay for general TSN flows without requirements on the traffic pattern. In general, the ATS shaper works well under a hybrid setting, e.g., the coexistence of both periodic and sporadic traffic. 

\begin{figure}[tb]
  \centering
  \includegraphics[width=0.7\columnwidth]{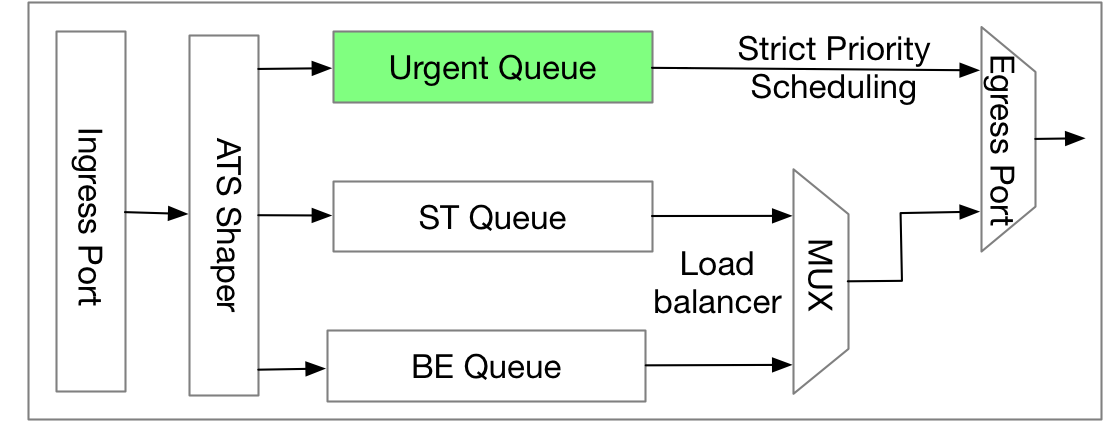}
  \caption{An ATS bridge example: ATS shaper determines the type of traffic at the ingress. ST represents the high-priority scheduled traffic, and BE represents the low-priority best-effort traffic~\cite{nasrallah2019performance}. }
  \label{Fig:ATS}
  %\vspace{-0.15in}
\end{figure}

Fig.~\ref{Fig:ATS} shows an example of an ATS shaper. The ATS shaper determines the traffic types at the ingress port for each incoming traffic. In the case of urgent traffic, it will be assigned to an urgent queue, which follows strict priority scheduling. For traditional high-priority scheduled traffic and low-priority best-effort queues, they follow a fair multiplexed transmission scheme.
%In addition, ATS can use the bandwidth effectively even when operating with mixed traffic loads, e.g., periodic and sporadic traffic, under a high link utilization.
 
%The ATS shaper was built on \textit{Urgency Based Scheduler (UBS)}~\cite{specht2016urgency} that operates according to two approaches: 1) Length-Rate Quotient (LRQ) and 2) Token-Based Emulation (TBE). %And the UBS scheduler builds on a packet service discipline, Rate-Controlled Service Disciplines (RCSDs)~\cite{zhang1994rate}. 
 %
%The UBS scheduler optimizes the RCSD disciplines on the following three aspects: 1) low and predictable worst-case delays guaranteed even at high link utilization rate; 2) low implementation complexity due to the separation of per-flow queues from per-flow states where flow state information is stored; 3) independence from the global time synchronization scheme. As such, the ATS shaper analyzes the each distinct traffic delay at each hop, and the overall delay from source to destination can be computed based on the network topology. 

%{\color{red} Need to revise... P17}

\begin{table*}[tb]
\small
\centering
\caption{Summary of different shapers.}
\begin{tabular}{|c|c|c|c|c|c|}
\hline
\textbf{Shaper}    & \textbf{Full name}                   & \textbf{Synchronization} & \textbf{Main Tech}            & \textbf{Topology dependence} & \textbf{Trigger} \\ \hline
TAS (Qbv) & Time-Aware Shaper           & Sync.           & TDMA                 & Dependent           & Cycle   \\ \hline
CBS (Qav) & Credit-based Shaper         & /               & Credit-based Shaping & Dependent           & Cycle   \\ \hline
PS (Qch)  & Peristaltic Shaper          & Sync.           & Double Buffering     & Independent        & Cycle   \\ \hline
ATS (Qcr) & Asynchronous Traffic Shaper & Async.          & Event-Trigger        & Dependent           & Event   \\ \hline
\end{tabular}
\label{Tab:Shaper}
%\vspace{-0.15in}
\end{table*}

Table~\ref{Tab:Shaper} provides a summary of different types of shapers in these standards. In the table, `Synchronization' represents the network model, which can be either synchronous or asynchronous, and `$/$' indicates that it does not require time synchronization. `Main Tech' refers to the main technology the shaper uses, e.g., TDMA. %\textit{Configuration} represents whether the schedule is based on the centralized model, e.g., YANG model, and the option \textit{offline} means it is based on a central scheduler while the option \textit{online} means it is based on the local information. 
`Topology Dependence' indicates whether the e2e latency is influenced by the adopted network topology, meaning that a larger topology with longer paths generally results in larger e2e delays, which also holds for the peristaltic shaper. `Trigger' represents the triggering mechanisms of the shaper, e.g., cycle, credit, or event.

\subsection{Reliability}\label{ssec:reliability}
{\rev Ultra-high reliability is another fundamental QoS requirement for industrial critical traffic. To achieve this, TSN provides several mechanisms to exploit the spatial redundancy of the communication channel and transmit replicated frames through multiple channels to tolerate both permanent and temporary faults~\cite{alvarez2017towards}. For this purpose, several standards have been defined in TSN, including IEEE 802.1CB and IEEE 802.1Qca. 
}
%TSN provides mechanisms for exploiting spatial redundancy of the channel %Nevertheless, the communication channel is especially vulnerable to transmission faults due to electromagnetic interference. One of the main objectives of TSN is to increase the reliability of communication by means of spatial redundancy. For this aim, several standards have been proposed, including IEEE 802.1CB and IEEE 802.1Qca. 
The IEEE 802.1CB standard manages creating and eliminating frame replicas to be transmitted through the existing path(s), while IEEE 802.1Qca allows for creating and managing multiple paths between any pair of nodes in the network. Besides, the IEEE 802.1Qci standard defines frame filtering and policing operations. 

%IEEE 802.1Qca describes new services to allow for the creation of multiple and non-shortest paths between nodes and further reserve resources for those paths. While IEEE 802.1CB manages the replication of streams, one frame duplicate will be transmitted through each one of these multiple paths. It also defines how to identify the stream that is replicated, and how frames should be replicated at the transmission and identified at the reception. The replication and elimination mechanisms will be implement on every element in the replicated paths, bridges and nodes. 

\subsubsection{IEEE 802.1CB Frame Replication and Elimination for Reliability (FRER)}

%\begin{figure*}
%  \centering
%  \includegraphics[width=0.8\textwidth]{Figs/CBexamp.png}
%  \caption{{\gw An example of IEEE 802.1CB on critical streams. It consists of a talker, listener, and relay system. Different colors among devices represent different flows~\cite{TSNCB}. }} 
%  \label{Fig:CB}
%  \vspace{-0.1in}
%\end{figure*}

The IEEE 802.1CB standard~\cite{8091139} lowers packet loss probability by replicating transmitted packets, sending them on disjoint network paths, and reassembling replicas at the receiver. IEEE 802.1CB is a self-contained standard that guarantees reliable and robust communication among applications through proactive measures to tolerate frame losses. 
Specifically, IEEE 802.1CB includes features such as sequence numbering, replication of each packet in the source station and/or network relay components, transmission of duplicates across separate paths, and elimination of duplicates at the destination and/or other relay components. By sending duplicate copies of critical traffic across disjoint network paths, IEEE 802.1CB minimizes the impact of congestion and failures, such as cable breakdowns. For instance, if one copy is lost during transmission, its duplicate still has a chance to be successfully transmitted. This scheme offers proactive redundancy for frame transmission but incurs additional costs on network resources. 

IEEE 802.1CB defines schemes for identifying packets belonging to streams and distinguishing them from other packets. Generally, to reduce network congestion, the number of packets to be replicated can be limited based on the traffic class and the link quality of the path. Additionally, to ensure frame recovery during transmission, a sequence generator identifies the replicated frames so the destination station can determine which frames should be discarded and which should be passed on. IEEE 802.1CB eliminates duplicates based on the sequence numbers carried in the frames. To enhance robustness and cope with errors, such as those caused by a stuck transmitter repeatedly sending the same packet, a recovery function is defined to remove packets with repeated sequence numbers from the stuck transmitter.

%The aforementioned operations on frames, e.g., duplication, elimination, and routing, are non-trivial tasks and it typically requires a centralized entity to manage these operations.  To improve the performance, these protocols are typically required to work with other protocols, such as IEEE 802.1Qcc TSN configuration and IEEE 802.1Qca path control and reservation, to further reduce the latency. 

%{\gw Fig.~\ref{Fig:CB} shows an example of the operation of 802.1CB. It provides three major functions: identification and replication of frames (for redundant transmission), identification of duplication frames, and elimination of duplicate frames. As shown in the figure, the sequence number is generated and encoded into each packet at the talker side. Sequence recovery functions at the relay devices eliminate duplicate packets, and the non-duplicate packets are copied as net member stream (e.g., with the sequence numbers unchanged) at two intermediate points. On the listener side, the duplicates are eliminated. In this example, this configuration can tolerate all seven possible one-link failures, and tolerate 16 of 21 possible two-link failures~\cite{TSNCB}~\cite{xu2016failure}. }  

\subsubsection{IEEE 802.1Qca Path Control and Reservation (PCR)}
The IEEE 802.1Qca standard~\cite{7434544} builds on two schemes: the Type-Length-Value (TLV) extension and the IS-IS (Intermediate System to Intermediate System) protocol~\cite{rfc1142}. The TLV extension is based on the Link State Protocol (LSP) of IETF, while the IS-IS protocol is used to establish connections among stations along the transmission path. This enables the IS-IS protocol to control bridged networks, extending the capabilities of the shortest path bridging (SPB)~\cite{802.1Q-2014} to manage multiple routes on the network~\cite{unbehagen2016path}. By integrating control protocols, IEEE 802.1Qca achieves several benefits, such as explicit forwarding path control, bandwidth control, and redundancy management. 
IEEE 802.1Qca provides mechanisms for bandwidth allocation and improves redundancy through various methods, such as protection schemes based on multiple redundant trees, local protection for unicast data flows based on loop-free alternates, and restoration after topology changes (e.g., following a failure event). %However, the adoption of IEEE 802.1Qca in industrial automation is uncertain, as networks in these fields are often pre-engineered, particularly in safety-critical application scenarios.

\subsubsection{IEEE 802.1Qci Per-Stream Filtering and Policing (PSFT)}
{\rev The IEEE 802.1Qci standard~\cite{8064221} defines protocols and procedures for filtering, policing, and service class selection on a per-stream basis. Filtering and policing functions include stream filters, stream gates, and flow meters to determine whether each frame is allowed to pass through to the egress queue. By setting up filtering rules and monitoring the passing frames, the standard can perform mitigation actions if violations are detected. 
Thus, IEEE 802.1Qci provides QoS protection when multiple streams share the same egress queue of a switch, preventing interference among them~\cite{bello2019perspective}. In addition, it improves network security against DoS attacks by identifying and dropping unauthorized or malicious transmissions, enhancing network robustness.
}
%defines the general procedures and managed objects for the bridge/switch to perform some frame operations, such as frame counting, filtering, and policing. 
%It also defines the service class selection for a frame based on its specified data flow, as well as a schedule for the synchronized cyclic events. %The schemes for frame policing and filtering defined in 802.1Qci include detection and mitigation of the disruptive transmissions by other systems, improving the robustness of the network.
%
%Generally speaking, time-based ingress policing plays a significant factor in scheduled networks, since flows are strictly scheduled to exit in a specific time-cycle from one node and enter the next node. This potentially enforces a time management scheme that uses the time to monitor TSN flows from ingress port to a TSN network entity. 
%IEEE 802.1Qci can help avoid the traffic overload induced, for example,  by erroneous transmission due to DoS attacks or devices malfunctions, which can potentially enhance the robustness of the network. In addition, IEEE 802.1Qci can be considered as a safeguard to the software errors on devices or attacks from the adversary. 

\subsection{Resource Management} % or Flow Management
%Qcc, Qat, Qcp, CS
{\rev Resource management is another key aspect of TSN that ensures the efficient allocation and utilization of network resources to meet the stringent requirements of industrial applications. It involves various mechanisms and protocols to manage network bandwidth, prioritize traffic, and maintain QoS through the definition of several standards, including IEEE 802.1Qcp, IEEE 802.1Qcc, and IEEE 802.1CS. 
}

\subsubsection{IEEE 802.1Qcp YANG Data Model}
% YANG-ref
{\rev IEEE 802.1Qcp~\cite{8467507} defines a YANG (Yet Another Next Generation) data model, specifying a data modeling language~\cite{rfc6020} used to model configuration data and state data manipulated by network management protocols such as NETCONF~\cite{rfc6241} and RESTCONF~\cite{bierman2017restconf}. Using the YANG model, IEEE 802.1Qcp allows configuration and status reporting based on Unified Modeling Language (UML) to manage IEEE 802.1 bridge devices.
}

{\rev YANG models the hierarchical organization of data as a tree, with each node representing configuration data, state data, RPC (remote procedure call) operations, and notifications. A set of related data nodes are organized into a module, the primary building block of the YANG model. 
}
To simplify the maintenance and management of complex modules, each module can be further subdivided into submodules. Readers can refer to~\cite{rfc7950} for more details on the YANG specification. The industry-wide implementation of the YANG model provides a universal interface to integrate resource management across diverse devices and equipment to fulfill the TSN standards. Besides, the YANG model can work with other enhanced TSN specifications, e.g., IEEE 802.1AX (Link Aggregation), for security. 

\begin{figure*}
  \centering
  \includegraphics[width=0.9\textwidth]{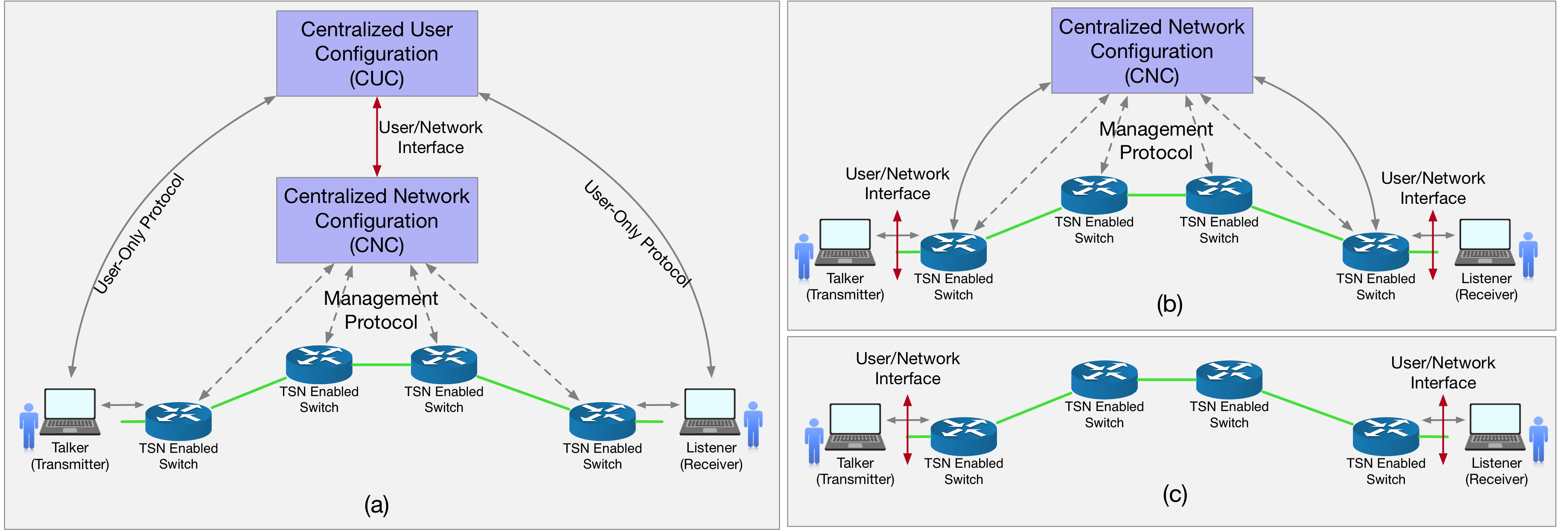}
  \caption{Three TSN resource management models: (a) fully centralized model; (b) centralized network / distributed user model; (c) fully distributed model.} 
  \label{Fig:YANG}
  %\vspace{-0.15in}
\end{figure*}

%Besides the above basic structures, the YANG model also defines the features of \textit{grouping}, \textit{augmentations}, \textit{constraints}, \textit{notifications} and \textit{operations}. Interested readers can refer to the YANG specification~\cite{rfc7950} for more details.

\subsubsection{IEEE 802.1Qcc SRP Enhancements and Performance Improvements}\label{sssec:qcc}
{\rev The IEEE 802.1Qcc standard~\cite{8514112} is an enhancement of the Stream Reservation Protocol (SRP) (IEEE 802.1Qat~\cite{5594972}) and deals with the configuration of TSN networks. IEEE 802.1Qat, originally designed for CBS shaper, manages the registration and reservation of resources within each bridge (e.g., buffers and queues) along the traffic path between the talker and the listener. Specifically, it serves as an admission control protocol where the talker registers the sending traffic with the required bandwidth, and it will be granted permission or not, depending on resource availability. This enables QoS management for streams with specific latency and bandwidth requirements. 
}

{\rev IEEE 802.1Qcc amends the IEEE 802.1Qat standard by extending the capabilities of SRP to adopt more complex shaping mechanisms, such as TAS with frame preemption. IEEE 802.1Qcc defines a user-network interface (UNI), which provides an abstract functionality between end stations (i.e., user side) and bridges (i.e., network side). The high-level idea is that the user specifies the requirement for the streams they want to transmit without knowing all the details about the network, and the network analyzes this requirement along with network capabilities and configures the bridges to meet the user requirements. 
} 
IEEE 802.1Qcc defines three configuration models~\cite{tian2019role}, as shown in Fig.~\ref{Fig:YANG}: the fully centralized model, the centralized network/distributed user model, and the fully distributed model. 
{\rev The fully centralized model introduces Centralized User Configuration (CUC) as the centralized manager for end users and provides the user requirements to the CNC through UNI. In the centralized network/distributed user model, the CNC configures TSN elements according to user requirements provided by the end bridges connecting end stations through UNI. In the fully distributed model, there is no centralized network configuration entity, and the network is configured in an fully distributed manner. }
Readers can refer to~\cite{gutierrez2017self} for more details.

\subsubsection{IEEE 802.1CS Link-Local Reservation Protocol (LRP)}
{\rev The IEEE 802.1CS standard~\cite{9416320} facilitates the replication of a registration database within a network link, i.e., from the device at one end to the device at the other end of the link. This enhances communication regarding resource registration among point-to-point devices and enables dynamic discovery, registration, and management of resources at a local level. The current 802.1Q Multiple Registration Protocol (MRP) supports databases up to 1500 bytes and significantly slows down when handling larger databases. To address this limitation, LRP is optimized to support the replication of registration databases on the order of 1 Mbyte. This enhancement enables new applications requiring much larger data sizes for configuration, registration, and reservation. LRP improves resource management efficiency since it operates within the local network segment without the need for centralized management. 
%LRP is a building block of the IEEE 802.1Qdd standard discussed below.
}

\subsubsection{IEEE 802.1Qdd Resource Allocation Protocol (RAP)}
{\rev IEEE 802.1Qdd defines RAP, which uses LRP from IEEE 802.1CS to support dynamic resource reservation for unicast and multicast streams in the fully distributed model. RAP also provides support for accurate latency calculation and reporting, and it is not limited to bridged networks. It aims to address issues present in the current IEEE 802.1Q Multiple Stream Reservation Protocol (MSRP), which has limitations in terms of the number of reservations, admissions, and configuration size in distributed stream reservation scenarios~\cite{osswald2021rap}. As of this writing, the standardization of RAP is still ongoing (IEEE P802.1Qdd Draft 0.9~\cite{Qdd}). 
%~\cite{osswald2021rap}
}
%aims to standardize a resource allocation protocol, which targets the distributed resource management and admission control protocol for networks with latency and bandwidth guarantees. 
%LRP protocol can also be implemented as a distributed TSN control model, namely, RAP protocol. The RAP protocol~\cite{chen2017resource} utilizes the LRP protocol to transmit the frame configuration information, which consists of the stream registration data and network resource reservation information.  %\sout{(3.5) Also, RAP can be used in a distributed scenario, which is more suitable for industrial automation applications~\cite{Qdd}}.
%This protocol can improve scalability (via LRP), support all TSN features, improve performance under high utilization, and enhance diagnostic capabilities, which is much suitable for industrial automation applications~\cite{nasrallah2018ultra}~\cite{Qdd}.

\vspace{0.1in}
{\rev \noindent \textbf{The advantages of TSN compared to existing industrial solutions.} 
After detailing the major capabilities of TSN, here we summarize its advantages over the existing Ethernet-based fieldbus systems. These advantages include openness, interoperability, convergence, and performance. First of all, openness and standardization are crucial to industrial automation since they promote wide cooperation among industrial partners. TSN is an open and standardized IEEE technology that is unaffiliated to any organization or company, and thus, the major manufacturers are very active in promoting TSN. Second, TSN ensures vendor-independent interoperability among the industrial devices, avoiding vendor lock-in and enabling system-wide connectivity. The combination of OPC UA and TSN, described in the following section, further fulfills the communication all the way from the sensor to the cloud. Moreover, TSN enables the convergence of IT and OT, which were previously kept separate in traditional industrial Ethernet-based protocols. Breaking down the communication barriers between IT and OT makes accessing data from industrial subsystems easier, where different traffic types can coexist in the network with their specific QoS requirements being met. In addition to the above advantages, TSN also excels in performance. While some advanced Ethernet-based protocols, e.g., PROFINET IRT, can also achieve deterministic real-time performance, TSN surpasses these solutions in latency (cycle time below 50 microseconds), jitter (less than $\pm$100 nanoseconds), and scalability (more than 10,000 network nodes)~\cite{bruckner2018opc}. Therefore, its openness, vendor-neutral interoperability, IT/OT integration support, and higher network performance, make TSN a highly effective and reliable choice for modern industrial automation. 
}

%% file: sec/Opportunities.tex
\section{Integrating TSN into Industrial Automation}
\label{Sec:oppo}
{\rev In this section, we first detail the key benefits of TSN for industrial automation and highlight the opportunities of integrating TSN into industrial automation through potential system-level integration.  We then elaborate on TSN traffic scheduling for achieving deterministic timing guarantees, a critical requirement for industrial automation applications. At last, as a crucial step before deploying TSN in real fields, we discuss the importance of TSN testbeds, highlighting their role in validating TSN performance in real-world industrial environments.
} 

% \subsection{Opportunities}
\subsection{Why Do We Need TSN in Industrial Automation?}

{\rev TSN is a game-changing technological advancement based on Ethernet and it is set to reshape the industrial communication landscape. This is mainly due to the many benefits offered by TSN to modern industrial automation networks, e.g., interoperability, convergence and determinism.}

{\rev As described in Section~\ref{sssec:trend}, the connectivity of industrial devices, i.e., interoperability, plays a critical role in industrial automation. At present, there are many tailored protocols and customized devices on the market for industrial Ethernet-based applications. While in many industrial application scenarios, customers may select different industrial Ethernet protocols to deploy their devices. This results in the protocol incompatibility and leads to vendor lock-in, which leaves the customers with only two options. One is to purchase all their devices from the same vendor even though some are not their best choices. The other option is to purchase their devices from multiple vendors but develop a convertible solution to integrate the devices, e.g., by implementing gateways to adapt among various industrial Ethernet protocols. However, both options are costly and can limit innovation on the factory floor~\cite{IntelWP}. 
Given the strength of TSN as an open IEEE standard, it guarantees compatibility at the network level among devices from different vendors. With TSN, a network consisting of multiple-vendor devices can inter-operate and be configured via a single standard interface. This provides customers with more options to build their system, avoids vendor lock-in, and enables connectivity across systems. 
The standardization technology carried out by TSN also enables the network structure to be standardized and flexibly extended without compatibility concerns. This also leads to a lower cost of ownership since the customers only need to replace existing switches with TSN switches, instead of duplicating networks and maintaining the additional hardware and software. 
%As a result, the automation industries have developed many hierarchical, purpose-built, and inflexible architectures~\cite{zhang2007hierarchical,wang2022harp,vitturi2019industrial}.
}

{\rev The IT/OT integration, accelerated by the rapid development of advanced manufacturing, acts as another critical enabler in the automation industry~\cite{schriegel2020ethernet}.
%accelerates the integration of Information Technology (IT) and Industrial Operation Technology (OT) in automation industry and further promotes the stringent demand for a unified network architecture~\cite{schriegel2020ethernet}. 
In legacy industrial Ethernet-based networks, different communication needs for IT and OT hinder the integration of these two fields. Specifically, larger bandwidth is typically required for data communication in the IT fields, while deterministic performance is the key for OT involving control operations. 
On the other hand, the digitization trend of industrial automation requires all types of data information (e.g., analog signals, sounds, images, and texts) must be converged. 
To this end, TSN provides the capability to break down communication barriers between various subsystems, including critical and non-critical systems. Different traffic types can coexist and be transmitted over the same network with no impact on traffic with a higher criticality level from traffic with lower priority. Network convergence provided by TSN makes it easier to access data from industrial systems and send them to the enterprise systems over standard Ethernet or the other way around without the need for gateways. 
}

{\rev Despite handling various traffic types across numerous devices in such converged networks, TSN can still provide deterministic performance guarantees, especially for critical traffic. 
%The current trends of automation and data exchange in industrial applications, i.e., Industry 4.0, are based upon digitization. That means analog signals, sounds, images, texts, and other information are converted into a computer-readable format for processing. For manufacturers to effectively use this information, data must be transmitted among numerous devices, e.g., sensors and equipment on the factory floor, and processed in real-time for informed decision-making by humans or other machines. 
%TSN provides standard Ethernet with the advantage of deterministic transmission, 
TSN ensures that the timing of critical traffic is predictable and consistent, which is essential for industrial automation applications. With deterministic message delivery, devices can communicate in real time, simplifying the configuration of systems, devices and applications, and increasing the productivity by enabling the machines to run cooperatively rather than independently. Informed decision-making by humans or other machines can also be processed in real time. This benefit of deterministic communication is achieved through TSN traffic scheduling based on network-wide time synchronization, which will be elaborated in Section~\ref{ssec:scheduling}.}

\subsection{{\rev TSN-based Converged Industrial Networks}}
{\rev TSN standardizes a set of technologies within the framework of IEEE 802.1 to provide guaranteed QoS. It is worth noting that TSN only resides at Layer 2 of the OSI model, i.e., it aims to provide bounded latency and jitter for point-to-point communication. Thus, TSN is not a complete communication protocol but rather can be taken as as a building block to provide the determinism foundation for converged industrial networks and it needs to be used in combination with higher-layer protocols to provide end-to-end QoS guarantee. On the other hand, industrial automation requires the Ethernet to support the convergence of all kinds of networks and traffic types typically found in an industrial setting. %TSN allows dissimilar industrial traffic to share the same network, and in such a converged network, data from various applications can be transferred simultaneously across a single line.
}

{\rev Converged networks in industrial settings require flexibility and scalability to use the same infrastructure (including small devices like sensor nodes, machine and production line control devices as well as big devices like data servers) for concurrent transmission of deterministic real-time communication (e.g., OT traffic) and non-deterministic best-effort communication (e.g., IT traffic). TSN is deemed as a key enabling technology to establish converged industrial networks with the following two trends: 1) Fieldbus\footnote{Here we refer to Ethernet-based fieldbus systems.} over TSN, and 2) OPC UA (Open Platform Communications Unified Architecture) over TSN~\cite{von2020tsn}. Table~\ref{Tab:converge} gives a summary of representative TSN-based converged industrial network solutions. Their details are described below.
}

\subsubsection{{\rev Fieldbus over TSN}}
{\rev At present, the industrial communication market is still dominated by Ethernet-based fieldbus systems~\cite{bruckner2019introduction} and there are many different fieldbus solutions in the market, e.g., PROFINET, EtherNet/IP, EtherCAT, Powerlink, and CC-Link. A major obstacle for today's Ethernet-based fieldbus systems is that they do not fulfill the convergence requirement of emerging industrial automation applications (e.g., a close IT/OT integration). Thus, combining industrial fieldbuses with TSN provides a way that can accomplish such requirement. 
There exist two main approaches for transmitting industrial fieldbus communication over TSN. One approach is to set up a new TSN network in accordance with every specification of the newly defined IEEE standards over Layer 1 and Layer 2 of OSI in factory networks so that fieldbuses can be transmitted without alternation. The other approach is to install active network gateways to convert all other network traffic between them to TSN-compatible Ethernet frames~\cite{von2020tsn}.   
}

{\rev Many fieldbus providers are already offering their products mapped to TSN, enabling seamless integration. For example, PROFINET over TSN~\cite{profinet} makes use of TSN features and supplements PROFINET on the Ethernet layer with IEEE standardized counterparts. With TSN, PROFINET is standing on a robust and future-proven foundation, which in turn creates more planning reliability for production and industrial solutions. On the other hand, existing PROFINET services (e.g. diagnostics and parameterization) and profiles (e.g. PROFIsafe, PROFIenergy, PROFIdrive) work as before on top of PROFINET over TSN and do not require any changes from the user.~\cite{schriegel2021migration} also suggests an improved migration strategy for PROFINET and designs a compatible bridging mode for TSN. 
}

{\rev EtherCAT over TSN~\cite{ethercat} defines a seamless adaptation to use both technologies and capitalize on their respective advantages without requiring any changes to the EtherCAT slaves. Adding EtherCAT segments as structuring elements in TSN reduces the complexity in backbones by using shared frames for a group of slaves and enabling internal configuration for a machine. TSN will protect EtherCAT segments from unwanted traffic while increasing the efficiency of the combined EtherCAT-TSN system. Combined EtherCAT and TSN can enhance flexibility at the automation cell level while maintaining total control of the various automation tasks. In~\cite{balakrishna2021simulation}, authors showcase the benefits of EtherCAT over TSN by using OMNeT++ to compare the cycle time and jitter under different scenarios of EtherCAT communications over a NeSTiNg TSN network. 
}

{\rev ODVA, which is a standards development organization and membership association, presents a recommended high-level approach for incorporating TSN capability into EtherNet/IP and identifies several major technical aspects of EtherNet/IP over TSN~\cite{ethernetiptsn}. 
TSN will be introduced in ODVA technologies as an optional and backward-compatible Data Link Layer for the EtherNet/IP implementation of CIP (Common Industrial Protocol). In~\cite{woods2017qos}, authors discuss specific use cases and examine how the TSN standards can be applied to EtherNet/IP networks to provide improved determinism and performance. 
}

{\rev CC-Link IE TSN~\cite{cclinkietsn}, developed by the CC-Link Partner Association (CLPA), is an open industrial network utilizing TSN  to seamlessly connect information systems to production sites. It uses Layer 3-7 of the OSI model, building on the TSN technology (Layer 2). With TSN, CC-Link IE TSN is able to increase openness while further strengthening performance and functionality. In addition to the above solutions with individual fieldbus systems,~\cite{chen2022tsn} designs a hybrid wired/wireless protocol conversion module that can realize intercommunication of three industrial Ethernet such as PROFINET, EtherCAT, and Ethernet/IP, and proposes a TSN-compatible frame to communicate with TSN based gateway. 
}

%{\rev Depending on the specific situation, devices with Ethernet-based fieldbus communication systems may be present or even dominant. Given that it is impossible that the existing industrial Ethernet protocols will vanish overnight, the above discussed fieldbus over TSN solutions make it possible for customers to transition to TSN easily and smoothly.}

\begin{table*}[tb]
\footnotesize
\centering
\caption{{\rev Summary of different TSN-based converged industrial networks.}}
\begin{tabular}{|c|c|c|c|c|}
\hline
\textbf{Organization or authors}        & \textbf{Type}  & \textbf{Year} & \textbf{Technology}      & \textbf{Summary}                            \\ \hline
PROFIBUS \& PROFINET International (PI)~\cite{profinet} & White paper    & 2021          & PROFINET                           & Principles, use cases, and architecture     \\ \hline
Schriegel et al.~\cite{schriegel2021migration}                        & Research paper & 2021          & PROFINET                           & Ethernet bridging mode                      \\ \hline
Karl Weber~\cite{ethercat}                              & White paper    & 2018          & EtherCAT                           & Integration approach                        \\ \hline
Balakrishna et al.~\cite{balakrishna2021simulation}                      & Research paper & 2021          & EtherCAT                           & Simulation-based case study                 \\ \hline
Woods et al. (ODVA)~\cite{woods2017qos}                     & Research paper & 2017          & EtherNet/IP                        & Use cases and challenges                    \\ \hline
Hantel et al. (ODVA)~\cite{ethernetiptsn}                    & Research paper & 2022          & EtherNet/IP                        & Technical recommendations                   \\ \hline
CC-Link Partner Association (CLPA)~\cite{cclinkietsn}      & White paper    & 2023          & CC-Link                            & Technical specification                     \\ \hline
%Chen et al.~\cite{chen2022tsn}                             & Research paper & 2022          & PROFINET, EtherCAT and Ethernet/IP & Architecture, protocol conversion mechanism \\ \hline
Li et al.~\cite{li2020practical}                               & Research paper & 2020          & OPC UA                             & Architecture and implementation             \\ \hline
Pfrommer et al.~\cite{pfrommer2018open}                          & Research paper & 2018          & OPC UA                             & Messaging mechanism and implementation      \\ \hline
Gogolev et al.~\cite{gogolev2018tsn}                          & Research paper & 2018          & OPC UA                             & Field device case study                     \\ \hline
\end{tabular}
\label{Tab:converge}
%\vspace{-0.15in}
\end{table*}

\subsubsection{{\rev OPC UA over TSN}}

{\rev Today's proprietary Ethernet-based fieldbus systems are broadly applied across different industrial automation networks to meet specific topology requirements, communication speeds, or latency guarantees. However, these communication protocols are often incompatible, resulting in fragmented networks that cannot seamlessly communicate with each other. OPC UA~\cite{leitner2006opc} was developed to solve this problem by allowing industrial devices operating with different protocols and on different platforms (e.g., Windows, Mac, or Linux) to communicate with each other. OPC UA supports two communication models, client-server (point-to-point communication based on TCP/IP) and publisher-subscribers (one-to-many communication supported by the new PubSub extension), without real-time capability. Thus, in conjunction with TSN, OPC UA over TSN under the pub/sub communication model allows deterministic transmission of real-time data and offers the flexibility and openness inherent to OPC UA~\cite{bruckner2019introduction,trifonov2023opc}. 
Note that, OPC UA over TSN and the above discussed fieldbus over TSN systems clearly overlap, but they are not replacing each other but will likely coexist for a long while. This is mainly due to the following fact. The strength of OPC UA, with real-time communication enabled by TSN, is that it allows different networks to communicate, especially at the factory- and enterprise-level. Industrial Ethernet, on the other hand, is primarily designed for communication between field devices and controllers. 
Below, we briefly discuss some OPC UA over TSN solutions. 
}

% TSN-based Converged Industrial Networks: Evolutionary Steps and Migration Paths
% Clock Synchronization in Future Industrial Networks: Applications, Challenges, and Directions

%\subsection{System-level Integration}
%{\rev TSN is widely used in industrial automation (e.g., manufacturing, oil and gas drilling, and transportation) not only because it provides deterministic performance guarantees but also due to its ability to integrate with many other technologies, e.g., OPC UA~\cite{leitner2006opc}, software-defined networking (SDN)~\cite{kreutz2014software}, and industrial 5G~\cite{aijaz2020private}. TSN advocates for integration rather than substitution, which is crucial for its application in industrial automation.}
%TSN can be applied to different industrial domains, e.g., manufacturing systems, and vehicle networks, and can combine with various technologies, e.g., OPC UA. 

{\rev \cite{li2020practical} proposes a communication architecture using the OPC UA and TSN for manufacturing systems. The proposed OPC UA TSN is a two-tier communication architecture, including the upper factory-edge tier and the lower edge-field tier. TSN is adopted as the communication backbone to connect different control subsystems in the field layer and the entities of the upper layers. OPC UA is adopted to realize horizontal and vertical information exchange between the entities of each layer. To validate the proposed OPC UA TSN, a laboratory-based experimental manufacturing system is implemented to demonstrate the feasibility and capability of the proposed architecture, as well as the real-time capabilities for industrial applications. 
\cite{pfrommer2018open} presents an OPC UA PubSub over TSN, which enables TSN to be used for the transport of OPC UA PubSub messages in practice. In the proposed approach, the message for the publisher is prepared in a (hardware-triggered) interrupt to ensure short delays and small jitter. Specific modifications are performed to allow the interaction between a best-effort standard OPC UA server and a real-time OPC UA PubSub publisher with access to a shared information model. The approach was implemented in open source based on the open62541 OPC UA SDK.
\cite{gogolev2018tsn} presents a case study on a TSN-enabled OPC UA integration for a field device. The evaluation indicates that the OPC UA integration of the field devices can be implemented using COTS software and hardware components. 
}

{\rev These R\&D efforts validate the potential of OPC UA TSN as a vendor-independent successor technology. OPC UA TSN is expected to quickly reveal itself as a game changer in the field of industrial automation, becoming the promising candidate to establish a holistic communication infrastructure from the sensor to the cloud~\cite{bruckner2018opc}. 
}

{\rev
\subsection{Traffic Scheduling}\label{ssec:scheduling}
Providing deterministic real-time performance, which is required by many mission-critical industrial applications, is one of the most promising TSN features. As described in Section~\ref{Sec:Sta}, the TSN Task Group has developed a suite of traffic shapers in the TSN standards, including TAS, CBS, PS, and ATS (see the summary in Table~\ref{Tab:Shaper}). These shapers provide a toolkit for managing network traffic to meet the diverse requirements of time-sensitive applications. Among these various shapers, TAS  stands out and draws special attention due to its ability to achieve deterministic timing guarantees by leveraging network-wide synchronization and time-triggered traffic scheduling mechanisms~\cite{zhao2022quantitative}, making it a key enabler to support deterministic real-time traffic in industrial automation. 

As described in Section~\ref{ssec:tsn}, a TSN switch is equipped with a set of time-gated queues to buffer frames from different traffic flows. The control of the queues (i.e., open or close) is specified by a predefined GCL with a limited number of entries, where each entry provides the status of associated queues over a particular duration. In addition, the priority filter in each switch utilizes a 3-bit Priority Code Point (PCP) field in the packet header to identify the stream priority and directs incoming traffic to the specific egress queue according to the priority-to-queue mapping. 
The configuration of GCL and traffic-to-queue mapping together define the network-wide schedule, which is determined by CNC and deployed on individual switches to guarantee the timing requirements of all time-triggered traffic. Traffic scheduling is thus one of the most critical problems in TSN and it results in a large amount of effort from both researchers and practitioners to develop various novel scheduling methods, especially for TAS shapers. 

Industrial applications that employ TSN as the communication fabric can be diverse regarding traffic patterns, network topology, deployment environment, and QoS requirements. Consequently, the specific TSN scheduling problem to be studied may vary significantly from the perspectives of the network model, traffic model, and scheduling model. 

$\bullet$ 
The \textit{network model} defines key attributes of the directed logical links in TSN, such as the propagation delay on Ethernet cables, processing delay on switches, link rate, number of available queues, and maximum GCL length. These parameters are typically determined by the capacity of the TSN switch or end station connected to each link.

$\bullet$
The \textit{traffic model} defines the parameters characterizing each TSN flow, including release time, period, payload size, deadline, and jitter. Each parameter can be individually modeled to capture the targeted traffic type based on specific industrial application scenarios. For example, the traffic model can be classified into fully scheduled traffic or partially schedulable traffic, depending on whether the release time of flows is predefined or determined by the corresponding talker. Additionally, based on jitter requirements, the traffic model can be categorized as a zero-jitter model or a jitter-allowed model.

$\bullet$
The \textit{scheduling model} specifies the constraints on the TSN systems under study, including queueing delay, scheduling entity, routing and scheduling co-design, fragmentation, and preemption. For instance, based on assumptions regarding queuing delay, scheduling models can be classified into no-wait and wait-allowed models. The scheduling entity determines whether the model is frame-based or window-based. Furthermore, depending on whether the routing path of each traffic flow is predefined or needs to be determined, scheduling models can be categorized as fixed routing models and joint routing and scheduling models.

Based on the above TSN model categorization, in a most recent TSN survey~\cite{xue2023real}, we present a systematic review and experimental study on 17 representative TAS-based TSN scheduling methods comparing their performance using various metrics\footnote{{\rev The established benchmark for performance evaluation of TSN scheduling methods is open-sourced. Please refer to our technical report~\cite{xue2023realtime} and GitHub repository~\cite{chuanyu2024}. We encourage the community to utilize this open-source toolkit to evaluate their scheduling methods to boost the development of TSN-related R\&D projects.}}. 
Unlike other TSN surveys (e.g.,~\cite{chahed2023tsn,stuber2023survey,deng2022survey,minaeva2021survey,seol2021timely,nasrallah2019tsn}) that either only provide a broad overview of the TSN standards or conduct conceptual comparisons of TSN scheduling methods,~\cite{xue2023real} offers comprehensive experimental comparisons among selected scheduling methods, including a diverse set of TSN system models and algorithms focusing on real-time scheduling of time-triggered traffic. The comparison results demonstrate that there is no one-size-fits-all scheduling method that can achieve dominating performance in all scenarios. Furthermore, diverse experimental settings complicate the fair evaluation of scheduling methods without introducing bias, which can make conclusions from previous studies only valid under specific settings. These findings also validate the inherent complexity of TSN traffic scheduling which is still an open problem. 
%More details on the experimental comparisons can be found in~\cite{xue2023real}. 
}

\subsection{{\rev TSN Testbeds}}
{\rev With all the benefits of TSN for industrial automation, before its deployment in real-world industrial sites, a crucial step is to validate its performance on ensuring all the stringent requirements posed by industrial automation applications. In general, three primary methods are used for evaluating TSN protocols and systems: theoretical analysis, simulation, and hardware testbeds~\cite{ulbricht2023tsn}. 
Many theoretical analysis frameworks have been developed to evaluate TSN, e.g.,~\cite{migge2018insights,guo2021formal,lv2020formal,zhang2019analysis}. However, these analysis frameworks make certain assumptions and abstract the behaviors of TSN systems compared to real-world settings. Simulation-based evaluation is another popular option and simulation tools, e.g., OMNeT++ and NS-3, have been widely used in TSN research~\cite{falk2019nesting,jiang2018time,debnath2023advanced,ozkaya2024simulating}. The advantages of simulations include flexibility, reduced cost, and scalability. However, they do not involve real hardware components, making it impossible to showcase the applicability in real industrial settings. Thus, a high-fidelity way is to use dedicated physical testbed based on real hardware to conduct well-defined experiments. 
}

{\rev Physical testbeds offer many benefits to the design and evaluation of TSN systems, enabling researchers and developers to explore, validate, and optimize their TSN solutions. The solutions can be rigorously evaluated in a controlled environment, ensuring that they meet the stringent industrial requirements. TSN testbeds also facilitate the assessment of interoperability between devices from different vendors. In addition, they help identify and address network configuration challenges and cybersecurity vulnerabilities, thereby mitigating deployment risks and ensuring a smooth transition to TSN-enabled industrial networks. 
However, the development of a TSN testbed is challenging from different points of view, ranging from implementation costs, sharing capability, and fidelity. Moreover, replicating real-world industrial conditions in a controlled testbed environment is difficult, and the cost and resource requirements, including specialized hardware, software, and skilled personnel, can be significant. 
}

{\rev Since TSN is a family of standards, TSN-related testbeds can be built to study different TSN aspects, including traffic scheduling, packet processing,  communication over-the-air, performance measurement, and network configuration. There have been a number of TSN testbeds developed for industrial applications and they can be generally classified into 1) general TSN testbeds, 2) OPC UA TSN testbeds, and 3) wireless TSN testbeds. General TSN testbeds (e.g.,~\cite{didier2017results,ulbricht2023tsn,quan2020opentsn}) focus on the fundamental TSN functions, e.g., scheduled traffic, credit based shaper, and time synchronization, to achieve real-time communication and deterministic behavior. OPC UA TSN testbeds (e.g.,~\cite{bruckner2018opc,schriegel2018investigation}) evaluate the integration of OPC UA and TSN to ensure the seamless flow of information among devices from multiple vendors. Wireless TSN testbeds (e.g.,~\cite{sudhakaran2021wireless,kehl2022prototype}) are built to explore the possibility of extending TSN capabilities to wireless media, including Wi-Fi and 5G. We will discuss the opportunities of wireless TSN in Section~\ref{ssec:wireless} and readers can refer to~\cite{zhang2024survey} for more details on the current TSN related testbeds. 
}

% OpenTSN: an open-source project for time-sensitive networking system development
%OpenTSN~\cite{quan2020opentsn} is an open-source project that supports rapid TSN system customization, which includes three features: SDN-based TSN network control mechanism, a time-sensitive management protocol, and a time-sensitive switching model. Different from the ASIC solutions targeted for TSN switches, OpenTSN provides FPGA-based hardware solutions for both TSN switches (TSNSwitch) and adapters (TSNNic), and a TSN network controller (TSNLight) for the control and management of a TSN network built-in OpenTSN. It opens all the hardware and software source codes so that designers can quickly and flexibly customize the TSN system according to their own needs. This can potentially maximize the reuse of existing code and reduce the customization complexity. Also, OpenTSN presents two FPGA-based prototypes with star and ring topology, and it shows that the synchronization precision of the entire testing network is under 32$ns$ and the transmission performance matches the theoretical analysis of the testing Cyclic Queue and Forwarding based TSN network. 

%Besides these efforts, there is a joint project of IEC SC65C/WG18 and IEEE 802 to define profiles for industrial automation. It defines TSN profiles for industrial automation, which select features, options configuration, defaults, protocols, and procedures of bridges, end stations, and LANs to build industrial automation networks~\cite{60802}. 

%% file: sec/Challenges.tex
\section{Challenges}
\label{Sec:Chal}
{\rev 
This section summarizes a number of challenges inherent to TSN standards that should be addressed. We follow the structure of Section~\ref{Sec:Sta} to discuss the specific challenges associated with each of the four pillars, i.e., time synchronization, latency guarantee, reliability, and resource management. 
}

\subsection{Time Synchronization}
{\rev Network-wide time synchronization is the foundation of all TSN features aimed at achieving deterministic real-time communication. IEEE 802.1AS is defined within TSN to provide accurate time synchronization using the gPTP protocol as described in Section~\ref{ssec:sync}. In the following, we discuss several key challenges that impact the accuracy and reliability of time synchronization, e.g., fault tolerance, synchronization overhead, and multi-level hierarchy.
}

%IEEE 802.1AS provides an accurate network-wide time synchronization mechanism, relying on the gPTP protocol. All gPTP components calculate both the residence time (e.g., processing time) and the link latency (e.g., propagation latency) based on their available own knowledge, and then exchange messages via the hierarchical master-slave structure, formed by the BMCA algorithm, to obtain the synchronized time.  
%In general, there are several key performance indicators to evaluate the performance of a time synchronization protocol. 
{\rev One of the primary challenges in TSN is to maintain precise synchronization across all network devices when applying the master-slave-based gPTP protocol. In a multi-hop TSN network, synchronization errors can occur, leading to synchronization failures~\cite{peng2023distributed}. These errors include time value error, i.e., incorrect time-related information (e.g., timestamp error) carried in propagated messages between nodes, and asymmetry in network delay, where the time difference between transmission delays from master to slave and vice versa causes errors~\cite{waldhauser2020time}. 
Clock drifts, due to the frequency drift of crystal oscillators, can cause gradual deviation of time clocks in various nodes over time, resulting in synchronization errors. In addition, security attacks, where compromised devices in the synchronization spanning tree propagate erroneous time information, can also lead to accumulated errors and synchronization failures. 
}

%The first performance indicator is the quality of local clock synchronization~\cite{flammini2010clock,gutierrez2017synchronization}. The quality largely depends on the underlying implementation of the network. 
%The second performance indicator is the reaction to potential security attacks. IEEE 1588, as the baseline protocol of IEEE 802.1AS, is subject to various security attacks. Examples of analyses in this area are~\cite{gaderer2006security} and~\cite{treytl2009security}, and IETF standardizes a threat model and security requirements for clock synchronization protocols~\cite{rfc7384}. In general, some traditional threats can be mitigated by the typical cyber-security procedures, such as encryption or cryptographic signatures, while other threats, such as delay attack~\cite{chen2022attack}, are difficult to mitigate by traditional approaches. For example, in a delay attack, a compromised bridge in the synchronization spanning tree may lie about its residence time and add values that are too long or too short to its response. Consequently, devices that receive this compromised response can suffer from a poor synchronization quality.  
%The third performance indicator relates to the ability of the protocol tolerate failures of involved devices and connections. Such failures may be a consequence of hardware faults or software bugs, and different industries may have different views on which failure semantics can be assumed for a failing device. 
{\rev To enhance resilience to synchronization failures, IEEE802.1AS only provides a basic level of redundancy, relying on BMCA (Best Master Clock Algorithm) to switch to a new Grandmaster (GM). To address this problem, IEEE P802.1ASdm~\cite{asdm} defines a hot standby mechanism to maintain two time domains simultaneously without relying on BMCA~\cite{yang2024clock}. }
While, addressing synchronization failures may require additional frequent message exchanges on timing information, consuming communication bandwidth and potentially causing back pressure on the centralized control plane, especially in large-scale applications~\cite{nasrallah2018ultra}. {\rev A trade-off between the synchronization accuracy and incurred overhead should be investigated where the settings of sync messages (e.g., transmission period) can be optimized. }
%One critical and open issue with regard to synchronization is that the frequent and periodic message exchange on the timing information would consume communication bandwidth and 
%For industrial cases, a centralized time synchronization system, e.g., software-defined network (SDN)-based design~\cite{li2015software}, can help release this situation with message exchanges only between a central synchronization controller and individual network entities. 
%In addition, the centralized scheme may be subject to single-point of failure attack.  
%IEEE 802.1AS/P802.1AS-Rev standardizes basic infrastructure mechanisms, such as distributing time from multiple grandmasters. However, it does not define how to construct a network to tolerate cases of synchronization failures. Besides, IEEE 802.1 AS and P802.1AS-Rev do not address the initial synchronization of the local clocks, e.g., the startup phase, or the system-wide restart phase. 

Moreover, industrial automation networks introduce further complexity with multi-level hierarchies on network switches, where different hierarchies may have varied synchronization quality. Since TSN standards operate at the MAC layer, even slight time slips in the upper layer can significantly affect the lower layer. The heterogeneity and accuracy differences among connected devices make a fully centralized time synchronization solution difficult to achieve in large-scale industrial automation. Therefore, applying a time synchronization scheme in industrial automation requires consideration of both network hierarchy and topology, which impacts the propagation mechanism of the synchronization messages. %\han{What role does the topology play here?}

\subsection{{\rev Latency Guarantee}}\label{ssec:latency}
{\rev In TSN, low latency guarantees are typically achieved through well-designed flow control, which includes traffic shaping and flow scheduling. Traffic shaping relies on various TSN shapers, each defining the traffic forwarding mechanism on TSN switches.  
Flow scheduling generates a network-wide schedule deployed on each device, specifying the timing of every transmitted frame. 
%Bounded low latency is typically achieved through flow control mechanisms, which set rules on how to efficiently and properly queue and forward the transmitted frames based on the specified schedules. 
%In current practice, most existing TSN flow scheduling mechanisms adopt a similar scheduling scheme where critical flows are commonly assigned some privileges to transmit, while other non-critical flows will be postponed. Most existing TSN schedulers and shapers provide transmission fairness, based on the assigned traffic class for the flows. 
%When multiple flows are transmitted with the same traffic class, the transmission selection algorithms choose the appropriate frame to transmit based on the available network resource and the ongoing traffic conditions.  
Building on the various TSN shapers introduced in Section~\ref{ssec:tsnlatency}, this section focuses on discussing the key challenges associated with each TSN shaper.}
%Here we only discuss the key challenges of several representative protocols, such as IEEE 802.1Qbv and 802.1Qbu. We then discuss several less used shaper schemes, which are the critical components in flow controls for industrial automation. 

\subsubsection{IEEE 802.1Qbv}\label{sssec:qbv}
{\rev Although the key idea of IEEE 802.1Qbv Time-Aware Shaper (TAS) mechanism is rather simple, there is an inherent complexity in generating the GCLs, i.e., deciding the right time instances to open and close the gates. This complexity is due to the NP-completeness of the TSN scheduling problem~\cite{lin2021queue} and thus no polynomial time scheduling algorithm exists unless $P=NP$. To this end, many TAS-based scheduling methods have been developed and these solutions can be classified into two categories. The first class aims to construct specialized search algorithms, i.e., by developing heuristics, meta-heuristics, or genetic algorithms (e.g., ant colony optimization (ACO)~\cite{ferrandi2010ant} and  meta-heuristics search algorithms~\cite{adly2013meta}). The second class leverages general-purpose tools, such as integer linear programming (ILP)~\cite{nayak2016time} or satisfiability modulo theories (SMT) solvers~\cite{robatmili2005optimizing,craciunas2014smt} to find the exact solutions. }

%Indeed, in general, the synthesis of communication schedules in real-world scenarios turns out to be an NP-complete problem, and to the best of our knowledge, no existing algorithm can generate the communication schedule in polynomial time~\cite{lin2021queue}. The synthesis of communication schedules can be differentiated into two types: the first type aims to construct specialized search algorithms, i.e., by developing heuristics, meta-heuristics, or genetic algorithms (e.g., ant colony optimization (ACO)~\cite{ferrandi2010ant} and  meta-heuristics search algorithms~\cite{adly2013meta}), and the second type leverages general-purpose tools, such as integer linear programming (ILP)~\cite{nayak2016time} or satisfiability modulo theories (SMT) solvers~\cite{robatmili2005optimizing}~\cite{craciunas2014smt} to construct the exact solutions. 
{\rev The primary challenge of generating TAS-based schedules is how to manage the trade-off between efficiency and precision. This trade-off arises from two main considerations. First, the choice of scheduling models -- such as whether to allow flow preemption, frame fragmentation, and whether to generate the schedule and routing path jointly -- impacts this balance. Using a more complex scheduling model, i.e., enabling the above options, can theoretically enhance system schedulability (i.e., the number of scheduled flows in the system) since it provides a larger search space. However, this also incurs higher computational overhead, which can be counterproductive in practice, especially in resource-constrained systems where a feasible schedule cannot be found by the algorithm in a reasonable amount of time. 
Another consideration for the trade-off is the choice of scheduling method category, i.e., heuristics or exact solutions. Specifically, heuristic algorithms demonstrate higher efficiency, particularly in large-scale networks, but they may not be able to find any feasible schedule in many cases. On the other hand, an exact algorithm can always find a feasible solution (if exists) to exhibit superior schedulability performance in small-scale networks.}

%{\rev Besides the challenge of schedule generation, the TAS shaper may introduce an extra sampling delay, since the TAS scheme requires continual sensing of the Ethernet channel to find the next available time slots. \han{I don't quite understand this challenge.} 
%The unsynchronized data flow can deteriorate the overall network performance. To obtain a lower latency, it requires the co-design of TSN end stations and gate scheduling on switches to schedule the e2e frame transmissions.  %Also, it is a nontrivial task to synchronize bridges even in a small sized network, since it might involve a fully centralized network to handle various situations. 
%Many commercial TSN switch products (e.g., TTTech Evaluation Board~\cite{tttechtsn} and Cisco Industrial Ethernet 4000 Switch~\cite{CiscoIE4000}) can support real-time and high-throughput (e.g., 1 Gbps) traffic with microseconds-level precision. However, the design of real-time TSN-compatible end station is much more challenging and remains an open problem~\cite{xue2024es,lehr2023design}. 
%}

{\rev %Besides the challenge of schedule generation, the TAS shaper imposes very high requirements on time synchronization since the unsynchronized data flow can deteriorate the overall network performance. 
Besides the precise configuration of switches, the TAS shaper imposes high performance requirements on end stations where it requires the co-design of TSN end stations and gate scheduling on switches to schedule the e2e frame transmissions. Many commercial TSN switch products (e.g., TTTech Evaluation Board~\cite{tttechtsn} and Cisco Industrial Ethernet 4000 Switch~\cite{CiscoIE4000}) can support real-time and high-throughput (e.g., 1 Gbps) traffic with microseconds-level precision. However, the design of real-time TSN-compatible end station is much more challenging and remains an open problem~\cite{xue2024es,lehr2023design}. 
Another notable challenge of TAS-based scheduling is the co-scheduling of time-triggered  (TT) traffic and synchronization traffic. If transmission collision between the two traffic types occur, it can cause synchronization error out of bound, resulting in network failure or deadline miss of TT traffic. }
%In general, the TAS shaper incurs much higher configuration complexity to achieve better performance, especially in a large scale application. The high configuration complexity in turn will limit the size of the target system.  %It is also critical to find the scheme to deal with the scalability issue, so that the system can be extended to be  large scale.  
%In general, scalability is typically a critical issue in industrial automation, which typically consists of a large number of network bridges. 
%\cite{xue2023real} presents a comprehensive experimental study on the existing TAS-based scheduling methods for TSN and provide foundational knowledge for the future studies on TSN real-time scheduling problems. 

\subsubsection{IEEE 802.1Qbu} 
IEEE 802.1Qbu Frame Preemption is beneficial to achieve bounded low latency, especially for critical traffic by preempting the transmission of non-critical traffic. {\rev The standard, however, only defines a one-level frame preemption paradigm where frames are classified into express frames or preemptable frames, depending on the criticality of the frames. While one-level preemption can ensure the transmission of high-priority critical traffic to some extent and is relatively simple to implement, it suffers from low flexibility since frames of the same category cannot preempt each other. To address this issue, some studies (e.g.,~\cite{ojewale2020multi,ojewale2021worst}) have proposed the concept of multi-level preemption. By introducing more frame categories, multi-level preemption allows for finer-grained preemption between frames. This approach enhances flexibility and can more effectively reduce frame latency. However, it also significantly increases the configuration complexity. For applications requiring deterministic real-time performance, the worst-case analysis of a multi-level preemption TSN network becomes highly complicated.}

{\rev TSN supports the concurrent operation of multiple shapers (e.g., TAS and CBS) on the same egress port, and thus utilizing frame preemption in such complex TSN setups can bring many benefits~\cite{debnath2024quantifying}. However, considering that the generation of the GCL is already an NP-hard problem as described in Section~\ref{sssec:qbv}, the use of frame preemption on combined TSN shapers would further elevate the difficulty and complexity of the configuration. Without highly effective and efficient traffic scheduling and configuration methods, combining so many functions could have adverse effects, such as incorrect configurations that fail to ensure timing correctness~\cite{ashjaei2022implications}. 
}

{\rev Since each occurrence of preemption divides the frame transmission into more segments, additional context switching is required. Therefore, the overhead introduced by preemption is another crucial consideration. Specifically, each preemption incurs a fixed overhead of 12 bytes, as well as the InterFrame Gap (IFG) of 12 bytes required between two consecutive transmissions~\cite{ojewale2018multi}. Moreover, when considering multi-level preemption, each preemption level introduces additional hardware implementation overheads. Thus, although the benefits of preemption are evident, addressing the trade-off between the performance gains from frame preemption and the associated overhead presents a significant challenge. 
}

\subsubsection{Other Shapers}
%Besides the TAS shaper discussed above, we now briefly discuss the shaping schemes in TSN flow control. 
{\rev The CBS shaper avoids starvation for best-effort flows at the expense of the transmission delay of higher priority and presumably more critical flows~\cite{benammar2018timing}. Although CBS is straightforward to implement, networks applying CBS are complex in analyzing the timing performance. In addition, TSN networks with high-volume traffic may suffer from poor performance under CBS in terms of delay guarantee~\cite{specht2016urgency}.
}
%Credit-based Shaper (CBS) is utilized in IEEE 802.1Qav, by incorporating the stream reservation protocol (SRP). The CBS shaper may induce a much higher delay, which can be up to 250$\mu s$ per hop which is not affordable in general mission-critical industrial control applications. %\sout{(3.16) To achieve a guaranteed latency in a TSN network, the CBS shaper may significantly reduce the utilization rate of network links. }  
%The PS shaper provides a mechanism that coordinates the operations on both enqueue and dequeue. In a PS shaper, all frames are required to be transmitted exactly within their time slots. Otherwise, the system might perform incorrectly. Thus, the PS shaper requires a straight alignment on the cycle times, which cannot be easily adapted to the asynchronous networks. 
The PS shaper coordinates operations for both enqueue and dequeue processes, ensuring that all frames are transmitted exactly within their designated time slots. This strict timing requirement means that PS shapers necessitate precise alignment of cycle times, making them less adaptable to asynchronous networks.
%The ATS shaper is based on Urgency-based scheduler (UBS)~\cite{specht2016urgency}. In UBS, the operations are typically performed on the per-hop basis, which is initially established on Rate-Controlled Service Disciplines (RCSDs). However, RCSD is subject to scalability issues. Specialized calendar queues can be used to address this issue, which require large memory pools. Also, it is not a trivial task to manage and control these large memory pools as the network size scales up. 
{\rev On the other hand, the ATS shaper aims to achieve bounded low latency for mixed-type traffic without global time synchronization. ATS provides less determinism for critical traffic than TAS but ensures a better average latency of all streams, as evaluated in~\cite{zhou2021simulating}. However, the current formula of ATS delay bound is rather conservative, where more precise timing analysis is required. 
}

{\rev While TSN defines various shapers that can provide real-time deterministic performance for critical traffic, this is usually based on the assumption of a homogeneous network where all devices support these shapers and there is global network time synchronization. However, industrial automation systems typically include a variety of devices, e.g., PLCs and other legacy equipment. TSN's vendor-independent interoperability feature allows for the existence of such heterogeneous networks within industrial systems. In heterogeneous networks with unscheduled and/or unsynchronized devices, meeting timing requirements remains a significant challenge. Designing effective scheduling mechanisms and timing analysis methods is essential to address this issue. These mechanisms need to ensure that even in the presence of diverse device capabilities and synchronization states, the network can still meet the stringent timing requirements of critical traffic~\cite{barzegaran2022real}.
}

\subsection{Reliability}
{\rev TSN enhances the reliability of industrial networks through several standardization efforts, including IEEE 802.1CB, IEEE 802.1Qca, and IEEE 802.1Qci, as described in Section~\ref{ssec:reliability}. However, these standards do not specify the exact implementation methods, leaving many research questions on fault tolerance to improve TSN reliability. In general, enhancing TSN reliability involves providing transmission redundancy, at both space and time dimensions. 
}

{\rev TSN standards typically use space redundancy. Specifically, IEEE 802.1Qca allows the creation of multiple paths between talkers and listeners for communication, while IEEE 802.1CB defines how to send duplicate traffic frames over different paths and eliminate redundant copies at the destination. This approach is well-suited for handling permanent faults, such as link breaks. The number of faults that can be tolerated depends on the number of redundant paths created~\cite{alvarez2019simulation}. However, space redundancy consumes significant network resources since the redundant paths are typically pre-established with bandwidth pre-allocated, regardless of whether faults occur during the  operation. In addition, configuring multiple redundant paths and frame copies increases the complexity of network scheduling. 
}

{\rev In contrast, time redundancy based on retransmission is more cost-effective. It creates multiple redundant copies of individual frames over time for retransmission. Unlike space redundancy, time redundancy is better suited for handling transient faults, e.g., packet loss and data error, which may result in incorrect reception and compromised data integrity~\cite{zhang2024fault}. The efficiency of time redundancy is also evident in its ability to differentiate the fault probabilities between different links. Indeed, the possibility of faults varies among links due to their physical characteristics. Therefore, time redundancy can allocate a different number of retransmissions for transmissions over different hops based on this information. Research in this area primarily focuses on how to meet reliability requirements, e.g., transmission success rates, with the minimum number of retransmissions~\cite{feng2023ret}. 
}

{\rev However, both space redundancy and time redundancy methods introduce additional network resource overhead, inevitably impacting other system performance, e.g., schedulability. To further improve resource utilization, adopting resource-sharing methods to provide redundancy is also effective~\cite{min2023effective,ma2023fault}. For example, in space redundancy methods, multiple paths can share one or more links, where partially disjoint paths can result in duplicate frames at intersection switches. In time redundancy methods, multiple traffic flows can share some time slots for retransmissions~\cite{zhang2022reliable}. However, these resource sharing methods must involve precise analysis of transmission success probabilities by considering various potential transmission scenarios, which posts great research challenge. An alternative approach to avoiding these highly complex analyses is to use learning-based methods, e.g., federated learning~\cite{feng2023ret}, to protect a network with probabilistic link failures. 
}

{\rev It is also crucial to make TSN resilient to adversarial attacks. TSN addresses this by defining IEEE 802.1Qci, which provides QoS protection through traffic suppression and blocking. 802.1Qci performs per-stream filtering and policing to protect against unnecessary bandwidth consumption, burst sizes, and malicious or improperly configured endpoints~\cite{xu2023recent}. It can also be used to confine network faults to specific areas, minimizing the impact on other parts of the network~\cite{luo2021security}. Although 802.1Qci is a published standard, there has been little research on deploying the standard on industrial network devices. One major challenge is how to configure the policing and filtering mechanisms of 802.1Qci, as misconfigurations can result in legitimate packets being filtered out or malicious packets being forwarded~\cite{ergencc2021security}, which degrades the network reliability and resilience. 
}

\subsection{Resource Management}
{\rev Resource management is essential for provisioning and managing network resources in TSN. It can significantly impact the network performance across various aspects, including network deployment, network configuration, traffic scheduling/routing, fault recovery, and network security. TSN primarily relies on the IEEE 802.1Qcc standard for resource management, complemented by the YANG model defined in IEEE 802.1Qcp, which provides a unified data template for network device configuration.
}

{\rev 802.1Qcc provides a set of tools for globally managing and reconfiguring the network, specifying three configuration models with regards to their architecture, as described in Section~\ref{sssec:qcc}. In general, each model\footnote{{\rev Since both the fully centralized model and the centralized network/distributed user model utilize CNC to configure TSN elements, we refer to them as the centralized model.}} has its strengths and weaknesses, and no single model is applicable to all industrial scenarios~\cite{shi2023recent}. 
The centralized model controls and manages traffic flows across the entire network, offering precise configuration and reconfiguration to meet timing and reliability requirements due to its global network knowledge~\cite{zhang2021dynamic}. However, this model has several flaws. The reliance on a single centralized controller makes the network vulnerable; if the controller fails, the network must maintain its current configuration and operating status until the controller is restored, rendering it unable to respond to network dynamics (e.g., adding new traffic) or failures. In addition, centralized models suffer from poor scalability. In large-scale networks, their response times can be considerably large due to reliance on the CNC and multicast broadcasting mechanisms to handle various network dynamics~\cite{zhang2019fully}. Furthermore, since a large amount of the computational workload is concentrated on the centralized controller, its computational performance can become a bottleneck for the entire network. 
On the other hand, the distributed model avoids the added complexity and single point of failure associated with centralized management and provides a much faster response to network dynamics since it does not require extensive configuration information exchange across the entire network. However, compared to centralized methods, it has slow network convergence and may result in transmission collisions, thus falling short of the network performance compared to those achieved by centralized methods. 
Therefore, selecting the appropriate resource management model and specific configuration methods based on the particular industrial application scenario and the corresponding application QoS requirements is a significant challenge. This decision must balance the trade-offs between complexity, responsiveness, scalability, and performance to ensure optimal network operation tailored to the unique demands of each industrial setting. 
}

{\rev Although IEEE 802.1Qcc is a published standard, the specified functions of the introduced CNC and CUC are not clearly defined. The implementation of the communication interface UNI between these TSN elements also needs further study. To this end, an ongoing standard, IEEE P802.1Qdj~\cite{10542670}, specifies enhancements to the UNI to include new capabilities to support bridges and end stations to extend the configuration capability. It also clarifies the functions of CNC and CUC, and stipulates the YANG model used for the communication between CNC and CUC. However, there is very limited research on these standards, leaving many challenging issues to be studied, e.g., the selection of appropriate resource management protocol among many candidates, including NETCONF, CORECONF, and RESTCONF~\cite{bhat2023coreconf}.
}

{\rev Furthermore, enabling efficient and effective network reconfiguration in response to various TSN network dynamics is a challenging task. For efficiency, industrial automation requires on-the-fly control and configuration to handle network dynamics without causing system downtime~\cite{chahed2021software}. This requires to avoid complex reconfiguration algorithms, e.g., SMT-based solutions, which require a long time to solve. 
For effectiveness, online reconfiguration must still meet stringent QoS requirements, particularly timing guarantees for critical traffic, even during dynamic adjustments. In this regard, centralized methods have their advantage since they have global network information. However, given the complexity of GCL configuration and routing determination, this remains a highly challenging problem.
}

{\rev Industrial automation systems may involve legacy or of-the-shelf end systems (e.g., PLC) that are unscheduled and/or unsynchronized. Dynamic reconfiguration for such heterogeneous TSN networks introduces another level of complexity since the TSN flows need to pass through the non-TSN network~\cite{reusch2023configuration}. This brings significant uncertainty to latency and jitter, requiring precise timing analysis to preserve the determinism of critical flows.
}

%% file: sec/Vision.tex
\section{Research Directions}\label{Sec:Visi}
In this section, we discuss several future research directions of TSN, including real-world field deployment, large-scale industrial network design, and wireless TSN. We believe that R\&D efforts in these areas will further support the seamless integration of TSN into industrial automation. 

\subsection{{\rev TSN Deployment}}
The TSN standards are still work in progress and require substantial modification, testing, and validation before wide deployment in real fields. In the following, we discuss several open R\&D problems related to TSN deployment and outline the future directions.

\subsubsection{Configuration Synthesis}
%From the functional perspective, there are several critical research directions of legacy TSN to worth exploring, e.g., configuration synthesis and security.
{\rev Given the network configuration and application requirements, the system designer needs to solve the so-called network-wide configuration synthesis problem~\cite{el2017network}, i.e., determining the set of combined mechanisms that can satisfy the application requirements. 
}
%\paragraph{Configuration Synthesis}
%Configuration synthesis addresses the issue of mechanism selections, that is, the decisions regrading the use or combination of mechanisms need to achieve the required features. %Also, configuration synthesis needs to find novel solutions to the scalability issues during the configuration. %Besides, the generation of TSN configurations can combine with other requirements, such as real-time and/or fault-tolerance, in configuration synthesis. 
Configuration synthesis is critical for industrial automation as different applications may have specific functional requirements. To maximize the benefits of applying TSN in the automation industry, the system designer must clearly understand the required functionality and make trade-offs in selecting specific TSN standards. 
{\rev The effects of using various standards in combination can lead to complex network configurations, potentially hindering the full utilization of TSN capabilities in industrial automation systems. This may further introduce extra costs during the product's lifetime if the selected technology needs replacement during or after deployment. Changing the selected standards would require significant redesign, installation, and re-verification~\cite{hallmans2020analysis}.}

\subsubsection{{\rev Coexistence of Shapers}}
{\rev With the advancement of industrial automation, many emerging industrial applications often have diverse QoS requirements. This requires TSN to support a range of time-sensitive applications by combining different shapers. This motivates an important future research direction to study the benefits and pitfalls of the coexistence of different types of shapers in the system. Some studies have already explored shaper combinations such as TAS + CBS (e.g.,~\cite{bezerra2022machine,houtan2024bandwidth}) and TAS + CQF (e.g.,~\cite{zhang2020coordinated,nie2024hybrid,wang2022hybrid}). When multiple shapers coexist in a system, they may interact with each other, potentially affecting overall performance. How to ensure that the key characteristics of TSN, especially e2e timing analysis, are maintained under these conditions deserves further investigation.
}

\subsubsection{{\rev Dynamic Reconfiguration}}
{\rev Industrial applications may suffer from unexpected dynamics (e.g., network topology updates and traffic specification changes) during the network operation. This requires dynamic TSN reconfiguration by adding, removing, or changing network devices and application tasks flexibly at run time. Although offline TSN configuration enables precise construction of communication schedules to provide deterministic performance for real-time industrial applications, it does not allow flexible network reconfiguration. To enable efficient and effective online reconfiguration, it requires a deep understanding of the dynamic configuration process, especially the associated timing overhead in each reconfiguration~\cite{pahlevan2019evaluation}. Then, effective dynamic reconfiguration methods based on different mechanisms (e.g., incremental reconfiguration~\cite{gartner2023fast} or pre-allocated partition~\cite{wang2021apas}) should be further explored. 
}

\subsubsection{Security}
{\rev Security is always a critical concern in industrial automation, and ensuring TSN security remains an open research topic. The IEEE 802.1 Security TG, part of the IEEE 802.1 WG, is actively working on enhancing TSN's secure capabilities, with ongoing cooperation between the IEEE 802.1 TSN TG and the IEEE 802.1 Security TG. However, as the automation industry becomes more open to the public, TSN-enabled systems will be exposed to various existing and novel attacks. Further research in TSN security is highly needed for early detection of these threats and development of effective mitigation strategies~\cite{bello2019perspective}. 
}

\subsection{{\rev Large-Scale Industrial Networks}}
{\rev In the current practice, TSN is mainly deployed in relatively small-scale LANs, enabling the connection among floor shop devices in factory-size networks. The maximum e2e latency of time-sensitive traffic classes can only be guaranteed up to seven hops, which significantly limits TSN’s scalability~\cite{tian2024large}. 
}

\subsubsection{DetNet}
{\rev To improve the scalability of TSN, the IETF DetNet group is working in collaboration with the TSN TG to develop standardization of IP layer deterministic forwarding services applied to Layer 3 routed segments. TSN/DetNet integration facilitates transforming isolated local real-time networks into integrated large-scale networks. Although DetNet standards are still under development, extensive research (e.g.,~\cite{wusteney2022analyzing,varga2023robustness}) has been conducted based on Request for Comments (RFC) documents~\cite{varga2021deterministic} and technical guidance drafts. However, research on DetNet over TSN is still at its initial stage, especially for deployment in large-scale industrial networks spanning large geographic areas. Ensuring consistent QoS performance (espeically for the timing guarantees) for such cross-network real-time communication poses many challenges. For example, long propagation delays between adjacent switches along a multi-hop path in a large-scale network can introduce significant jitter and reduce synchronization precision. Additionally, traffic scheduling in a cross-network setting becomes more complex as relying on a centralized controller (i.e., CNC) to pre-compute the network-wide schedule is not feasible anymore. Exploring distributed (e.g.,~\cite{shen2022distributed}) or hierarchical scheduling mechanisms (e.g.,~\cite{wang2022harp}) could lead to be possible solutions. 
}

\subsubsection{Virtualization}
{\rev A large-scale industrial automation system is typically an integration of heterogeneous computing and communication platforms containing diverse hardware, e.g., multi-core CPUs, GPUs, MCUs, and FPGAs. The stringent timing requirements further drives the industrial automation systems to employ the edge-cloud computing paradigm with a hierarchy of computing resources. To manage these heterogeneous resources, resource virtualization is an enabling technique that can help reduce the operation expenses and increase the system flexibility and scalability since applications running on virtual machines (VMs) can be easily managed (e.g., create, migrate or delete)~\cite{wang2023resource}. However, the use of TSN in virtual environments is a relatively new trend as the TSN standards were originally intended for bare-metal industrial applications and recently there have been some pioneering work on this topic (e.g., architecture hypotheses~\cite{leonardi2020towards} and testbed validation~\cite{garbugli2022framework}). Despite the potential advantages provided by resource virtualization, it is still an open research problem with many challenges unsolved. First, virtualization may introduce a source of unpredictability (e.g., unpredictable latency caused by VMs running on adjacent cores) that may lead to the loss of determinism. To achieve the desired flexibility, VM placement and dynamic VM migration (e.g., virtual PLCs) pose challenges in online TSN scheduling in response to dynamic changes of application requirements. In addition, to mitigate any form of overhead, lightweight virtualization techniques have become the standard technology for edge components, e.g., using containerization instead of hypervisor-based VMs~\cite{garbugli2023kubernetsn}. The highly distributed nature of edge cloud applications is a challenge to effectively supporting the most performance-demanding components in containerization frameworks. 
}

\subsection{Wireless TSN}\label{ssec:wireless}
Most existing industrial automation systems rely on Ethernet-based fieldbus communication, which are based on wired connections. Applying wireless technologies to the automation industry provides many obvious advantages, e.g., reduced wiring cost and improved device mobility. 
{\rev Many industrial automation use cases can directly benefit from TSN capabilities over wireless, e.g., closed loop control, mobile robots, and autonomous ground vehicles~\cite{cavalcanti2020wireless}. However, given the inherently unreliable characteristic of wireless connection, achieving wireless TSN is challenging~\cite{cavalcanti2019extending}, particularly in providing deterministic timing and reliability guarantees. Wireless media has fundamental differences from their wired counterparts, e.g., varied transmission capacity depending on link quality and unreliable nature due to stochastic properties of the channel and interference. These challenges motivate a number of future research topics. 
}

\subsubsection{{\rev Time Synchronization}}

{\rev Both industry and academia have been actively working on the design and development of wireless TSN, where IEEE 802.11 and 5G are considered the two major candidates. For this aim, achieving accurate time synchronization is the first step towards making TSN available on wireless networks, and it is the foundation for time-critical traffic scheduling to achieve deterministic real-time communication. Different from wired industrial networks, time synchronization over wireless networks needs to tackle several challenges (e.g., high delay variation and imprecise timestamping), and there is a rich literature on analyzing or providing real-world implementations of the integration of wired and wireless clock synchronization for both IEEE 802.11 and 5G. 

For IEEE 802.11, there are mainly three messaging schemes to perform clock synchronization: 1) IEEE 802.1AS messaging relying on the de facto PTP standard~\cite{val2021ieee,gundall2021integration}, 2) IEEE 802.11 messaging by integrating Fine Timing Measurement (FTM) into 802.1AS~\cite{thi2022ieee}, and 3) low-overhead beacon-based time synchronization mechanism~\cite{romanov2020precise,haxhibeqiri2021enabling}. 
}

{\rev For 5G, the clock synchronization support is standardized in the Third Generation Partnership Project (3GPP) Release 16~\cite{3gpp2019release16} and mainly two time synchronization approaches are considered~\cite{striffler2019time}: boundary clock and transparent clock. The former requires the 5G Radio Access Network (RAN) to have a direct connection to the TSN master clock based on IEEE 802.1AS~\cite{shi2021evaluating}. The latter is achieved via PTP messages among any forwarding devices by passing relevant time event messages~\cite{striffler20215g,muslim2024synchronizing}. While the boundary clock approach is simpler to implement, the transparent clock approach is mostly preferred due to its much higher accuracy~\cite{satka2023comprehensive}. 
Despite significant research progress, time synchronization for wireless TSN still faces many challenges deserving further investigation, including the lack of hardware-timestamping, synchronization errors during handover, and asymmetry in uplink/downlink propagation delay which adversely affect the synchronization process. 
}

\subsubsection{{\rev Traffic Scheduling}}
{\rev To meet deterministic timing guarantees in wireless TSN, besides precise time synchronization, another critical research area is timing-aware traffic scheduling. IEEE 802.11's default medium access is contention-based and non-deterministic. Thus, a significant amount of research has explored replacing/improving traditional 802.11 MAC with TDMA-based MAC protocols (e.g.,~\cite{wei2013rt,seijo2020w,yun2022rt}). Other efforts have focused on implementing 802.1Qbv on the network stack using TSN functionalities and tools available in the Linux kernel (e.g.,~\cite{sudhakaran2022wireless}). %, but they cannot guarantee the required performance when used with the 802.11 MAC layer~\cite{avallone2023controlled}. 
In the meantime, IEEE 802.11 is rapidly evolving to support time-sensitive applications in industrial automation. For example, Wi-Fi 6 (802.11ax) supports several methods (e.g., the scheduled trigger frame (TF)-based access scheme) to enable wireless TSN-capable access points (APs) and to ensure nearly deterministic transmissions. Further enhancement to the 802.11ax TF is also under consideration by Wi-Fi 7 (802.11be) to deterministically schedule 802.11 frames~\cite{adame2021time}. It is expected that more research will emerge to address other open challenges, e.g., supporting ultra-low latency and frame preemption~\cite{cavalcanti2022wifi}. 
}

{\rev 5G, as another wireless TSN candidate, does not share the same IEEE 802-based link layer as Ethernet and Wi-Fi, while 5G-TSN integration is also feasible via translation interfaces defined in 3GPP Rel. 16~\cite{3gpp2019release16}. 5G can be integrated within the TSN network as a logical TSN bridge where the 5G core and RAN remain hidden from the TSN network. To inter-operate between TSN and 5G systems, 3GPP introduces the TSN translator functionality at the interconnection points between both networks. The translator functionality, both in the device side and the network side acting as TSN ingress and egress ports, is to configure all parameters necessary to coordinate 5G and TSN~\cite{welte2024tsn}. These translators realize the configuration of the 5G system in order to fulfill the required TSN deterministic transmissions with bounded latency. 5G ultra-reliable low-latency communications (URLLC) provide a good match to TSN features by enabling increased reliability and latency below 1 $ms$. Significant research works (e.g.,~\cite{zhang2023real,zhang2023contention,pan2024multi,ginthor2019analysis,shen2022qos}) have also studied the real-time scheduling problems of URLLC traffic in industrial applications to meet their stringent timing requirements. These solutions, however, are more suitable in standalone industrial 5G networks instead of 5G-TSN integration systems which must follow the schedule specified by CNC in TSN. To achieve this, internal configuration is required for the 5G system, including mapping traffic classes in TSN into a predefined 5G QoS indicator (5QI) and leveraging hold \& forward buffering mechanism which is identical to the gate scheduling behavior of TSN GCL~\cite{larranaga2020analysis}. Although 3GPP specification provides a comprehensive mapping from 5G to TSN traffic shaping and scheduling, the wireless nature that allows mobility and frequent changes in the network layout requires further enhancements to the traffic scheduling mechanism design.
}

\subsubsection{{\rev Reliability}}
{\rev In addition to time synchronization and traffic scheduling, guaranteeing the reliability of transmissions is another key challenge to enable wireless TSN. The ultra-reliability feature in wireless networks is typically pursued through enabling transmission redundancy in different manners including 1) intra-frame redundancy, 2) inter-frame redundancy, and 3) multi-path redundancy. Intra-frame redundancy introduces redundant bits within a frame to increase the probability of successful reception of a frame. 802.11 and 5G both support intra-frame redundancy via the configuration of modulation and coding scheme (MCS) specifying the ratio of redundant bits in a frame. Inter-frame redundancy performs frame retransmissions either actively (i.e., after detecting transmission failure through ACK) or passively (i.e., reserving multiple frame copies). Active redundancy is spectrum-efficient but suffers from higher transmission latency. Thus, passive redundancy is a compelling method to achieve ultra-reliability in wireless TSN without sacrificing latency. In multi-path redundancy, multiple copies of a frame are transmitted to the destination through different paths or links. 802.11 supports multi-link operation allowing a station to simultaneously maintain multiple 802.11 links across the 2.4, 5, and 6 GHz bands. 5G can enable multi-path redundancy through setting up redundant Protocol Data Unit (PDU) sessions where different solutions can be applied~\cite{atiq2021ieee}. 
Currently, the transmission interference is still a major hurdle to achieving ultra-high reliability, especially for communications in unlicensed bands like Wi-Fi. Power management is another direction since an increment of the transmission power improves the transmission reliability but may decrease the power efficiency of the wireless system. 
}
%Synchronization: IEEE 802.1AS (WiFi, 5G~\cite{schungel2020analysis,striffler20215g,muslim2024synchronizing}).

%5G serves as a good candidate for connecting devices, e.g., connecting industrial sensor/actuators wirelessly to a wired TSN network~\cite{hossain20155g}. Compared with 4G, 5G offers several new features, such as better reliability, lower transmit latency, and flexible deployment into the current infrastructure. %And, integrating 5G into the TSN network, as a wireless solution, provides many new insights, e.g., avoiding the limitations of cable installations. 

%Integrating wireless technologies such as 5G into the industrial TSN systems, requires several conservative considerations to guarantee the reliability of wireless TSN. For example, in wireless communication, jitter is a common case, and we need to reduce the impact of jitter. When integrating wireless TSN into wired systems, there are several considerations: wireless configuration of wired TSN; hybrid wired-wireless time synchronization; wireless TSN scheduling; wireless redundancy for wired TSN; and wireless TSN switch deployment~\cite{bush2018industrial}. The adopted wireless technologies should satisfy these considerations, and a centralized resource management, e.g., based on the Yang model, is mandatory. 

\subsubsection{{\rev Wireless Security}}
{\rev Providing security and safety guarantees is critical for industrial automation systems. TSN defines the 802.1Qci protocol to block malicious devices or attacks and 802.1Qci provides traffic filtering and policing schemes at the ingress port of switch to prevent unidentified traffic, thereby improving network security. Many researchers also discuss the design of fault detection methods and encryption mechanisms based on 802.1Qci (e.g.,~\cite{luo2021security,wang2023research}) to further enhance the network security. Many other strategies (e.g., authentication, encryption and decryption, intrusion prevention) may also be deployed to achieve e2e security in TSN. However, the trade-off between the cyber security and TSN performance must be considered since cyber security strategies can introduce additional traffic transmission delay which further impact the determinism of the network. %Another promising direction is the use of blockchain technology, which provides the confidentiality strategy of data and proposes the Hash algorithm to ensure the integrity of data~\cite{bodkhe2020blockchain}. 
}

{\rev Comparing Ethernet-based TSN networks with wireless TSN networks, they share similar security objectives at a high level, but wireless networks are more vulnerable to attacks, e.g., eavesdropping and tampering~\cite{nazir2021survey}. To address these security concerns, Wi-Fi Protected Access (WPA) is an authentication and key management protocol developed for encryption in Wi-Fi and WPA2 is retired by the new standard WPA3 to make Wi-Fi more secure~\cite{kwon2020evolution}. For 5G, 3GPP defines several security domains, e.g., network domain security, user domain security, and application domain security, with many solutions standardized throughout the evolution of cellular technologies, including mutual authentication and authorization of the network and the UE, integrity protection of the RRC-signaling and NAS-signaling, etc.~\cite{3gpp.33.501}. Additionally, in the context of 5G-TSN deployment, unique challenges such as clock skew in GM-based time synchronization, denial of service (DoS) attack, and rogue base station (RBS) should also be investigated~\cite{sethi2022security}. 
}

{\rev In summary, supporting wireless TSN requires careful selection of design approaches, considering several trade-offs in the design process. These include the trade-off between scheduling complexity and handover delay in dealing with user mobility~\cite{avila2024unlocking}, the trade-off between deterministic performance guarantee and associated radio resource costs~\cite{liu2023multicast}, and the trade-off between the reliability and traffic aggregation overhead~\cite{zhang2022enabling}. 
Based on the development of wireless TSN, the next research step is clear that wireless TSN and wired TSN must be integrated to create hybrid TSN networks~\cite{seijo2021tackling}. The integration of the technologies poses several challenges. Essentially, a hybrid TSN network must maintain the TSN features across the different communication domains and technologies, including guaranteed e2e latency, clock synchronization, and coexistence of traffic flows with different criticality requirements. 
}

%% file: sec/conclusion.tex
\section{Conclusion}
\label{Sec:Concl}
{\rev The industrial automation market is still dominated by Ethernet-based fieldbus systems, particularly those with real-time capabilities, e.g., EtherCAT, PROFINET IRT, POWERLINK, and SERCOS III. Although these technologies are based on conventional Ethernet, they are not designed to interoperate with field buses from other vendors. In the context of industrial automation, a large number of vendor-crossing devices with diverse QoS requirements are expected to communicate across all levels of the automation pyramid. Thus, TSN has the potential to enable modern industrial automation by establishing universal physical and data-link layer standards. 
}
TSN consists of a set of Ethernet-based protocols and standards designed to address a wide range of practical industrial use cases with guaranteed timing requirements in heterogeneous networks. TSN encompasses a broad scope, making it critical to understand the standards systematically rather than focusing on just one characteristic or component. 
This paper provides a comprehensive review of TSN standards in industrial automation, including both published standards and in-progress drafts. We specifically focus on the automation industry, discussing the challenges and opportunities when applying TSN to industrial control applications. In addition, we highlight promising research directions for TSN design and development in industrial automation, such as optimizing current TSN standards and integrating TSN with other technologies. 

%\section{Acknowledgement}
%The work is supported in part by the National Science Foundation Grant CNS-1932480, CNS-2008463, CCF-2028875, CNS-1925706, and the NASA STRI Resilient Extraterrestrial Habitats Institute (RETHi) under grant number 80NSSC19K1076.